\documentclass[aps,twocolumn,showpacs,nofootinbib]{revtex4}
\usepackage[dvips]{graphicx}
\usepackage{amsmath}
\newcommand{\bea}{\begin{eqnarray}}
\newcommand{\eea}{\end{eqnarray}}
\newcommand{\be}{\begin{equation}}
\newcommand{\ee}{\end{equation}}

\newcommand{\BSCCO}{{Bi$_2$Sr$_2$CaCu$_2$O$_{8+x}$ }}
\newcommand{\YBCO}{{YBa$_2$Cu$_3$O$_{7-\delta}$ }}

\newcommand{\rr}{{\bf r}}

\newcommand{\nn}{\nonumber}

\def\k{{\bf k}}

\def\q{{\bf q}}
\def\v{{\bf v}}

\begin{document}


\title{Antiferromagnetic correlations and  impurity  broadening of NMR linewidths in cuprate superconductors}
\author{J. W. Harter$^1$, B. M. Andersen$^{1,2}$, J. Bobroff$^3$,  M. Gabay$^3$, and P. J. Hirschfeld$^{1,3}$}
\affiliation{$^1$Department of Physics, University of Florida, PO
Box 118440, Gainesville FL 32611, USA \\ $^2$Laboratoire de Physique
Quantique, ESPCI, 10 Rue de Vauquelin, 75231 Paris, France \\
$^3$Laboratoire de Physique des Solides, Universit\'e Paris-Sud,
91405 Orsay, France }

\begin{abstract}
We study a model of a $d$-wave superconductor with strong
potential scatterers in the presence of antiferromagnetic
correlations and apply it to experimental nuclear magnetic
resonance (NMR) results on Zn impurities in the superconducting
state of \YBCO (YBCO).  We then focus on the contribution of
impurity-induced paramagnetic moments, with Hubbard correlations
in the host system accounted for in Hartree approximation. We show
that local magnetism around individual impurities broadens the
line, but quasiparticle interference between impurity states plays
an important role in smearing  out impurity satellite peaks.  The
model, together with estimates of vortex lattice effects, provides
a semi-quantitative description of the impurity concentration
dependence of the NMR line shape  in the superconducting  state,
and gives a qualitative description of the temperature dependence
of the line asymmetry. We argue that impurity-induced
paramagnetism and resonant local density of states effects are
both necessary to explain existing experiments.
\end{abstract}

\pacs{74.25.Ha,74.72.Bk,74.81.-g}

\date{\today}
\maketitle
\narrowtext
\section{Introduction}
The response of a correlated electron system to a local
perturbation can often provide important information about the
ground state of the pure system.  This principle has been
successfully applied to the substitution of various impurities,
particularly Zn, Ni, and Li for Cu in the CuO$_2$ planes of
various  high-T$_c$ superconductors, particularly in  \YBCO. In
addition to traditional studies of the effect of impurities on
bulk properties, local probes like NMR and scanning tunnelling
spectroscopy (STS) have provided considerable information about
how electronic wavefunctions are distorted near the impurity site.
In the normal state of this system, NMR has shown that impurities
enhance local antiferromagnetic
correlations\cite{Kataev,AVMahajan:1994}, and in the presence of
the applied DC field display a staggered pattern of magnetization
which decays over a few lattice spacings. This polarizability
$\delta \chi$ has moreover a characteristic temperature
dependence, which is Curie-like $\delta \chi \sim T^{-1}$ in the
underdoped system, but evolves to Curie-Weiss-like behavior
$\delta \chi \sim (T+\theta)^{-1}$ in the optimally-to-overdoped
range\cite{JBobroff:1999}.  Because $\theta$ increases rapidly
with doping, it has sometimes been interpreted as a Kondo
temperature, enhanced in the presence of higher carrier densities
capable of screening the magnetic moment induced by the impurity.
Other pictures of this phenomenon, which do not rely on Kondo
screening, have been put forward as well.  For example, in the
weak-coupling approaches of Bulut\cite{NBulut:2000,NBulut:2001}
and Ohashi\cite{YOhashi:2001}, an extended potential is found to
produce a Curie-Weiss-like local susceptibility in a Hubbard model
treated in mean field, due to the coupling of the
antiferromagnetic $\q=(\pi,\pi)$  response of the lattice system
to the uniform $\q=0$ response by the inhomogeneity.


In the superconducting state, interpretation of the NMR signal is
complicated by the intrinsic field distribution introduced by the
vortex lattice, and by the vanishing of the $\q=0$ susceptibility
in the singlet pair state.  On the other hand,
 Ohashi\cite{YOhashi:2002}
 argued that the mode-coupling effect induced by the inhomogeneity
 persists and is relatively enhanced by the opening of the gap.
Experimentally, the enhancement of the local susceptibility was
indeed found
 below $T_c$\cite{JBobroff:2001}, and  tended to  a large constant value
 at very low $T$.   Recently, Ouazi {\sl et al.}\cite{SOuazi:2006} measured the evolution of the $^{17}$O
 NMR
line with increasing Zn concentration, and observed the formation
of the staggered polarization cloud for the first time in the
superconducting state.  They argued that the primary line shift
was due to the nearly field-independent vortex distribution, and
that the broadening was a combination of the enhanced impurity
effect and the simultaneous narrowing of the vortex field
distribution due to the increased penetration depths $\lambda$ in
the dirtier systems.  Missing from this picture is an
understanding of the magnitude of  the impurity broadening and
how it really occurs; if one considers only a single impurity one
expects large values of the magnetization on the Zn nearest
neighbor sites, which should lead to a well-defined satellite line
as in the case of  NMR on Li impurities in the normal
state\cite{JBobroff:1999}.  These have not been detected in
samples with Zn concentrations at the per cent level.

An alternate picture of the observed phenomena in the
superconducting state is obtained if one considers the  local
 susceptibility due to quasiparticles in the $d$-wave superconductor,  proportional to the
 local density of quasiparticle states (LDOS) at the Fermi level.
 A significant
 enhancement of the local susceptibility is then to be expected from
 quasiparticle bound states alone.  Williams {\sl et al.}\cite{GVMWilliams:2000}
 proposed that these quasiparticle resonant states--corresponding to those  imaged by  STM
 experiments around Zn atoms in \BSCCO (BSCCO-2212)\cite{SHPan:2000}--might be entirely responsible for the
 enhanced magnetic response near Zn seen in NMR. Chang {\sl et
 al.}\cite{JChang:2004} then argued that for a single nonmagnetic impurity,  the temperature dependence of the
 observed spin-lattice relaxation time and Knight shift could
 indeed be qualitatively understood in terms of LDOS enhancement due to impurity bound states alone.

There are some difficulties with the naive interpretation of the NMR
measurements entirely in terms of the LDOS enhancement near
impurities, however.  First of all, significant $T$-dependent
enhancements of local susceptibilities near nonmagnetic impurities
occur in the normal state of optimally doped cuprates as well.  An
``LDOS-only" approach cannot account for this since impurities do
not produce LDOS resonances in the normal (metallic) state.
Secondly, the NMR experiments on these materials clearly show that
the magnetization near a Zn alternates in sign.
This is incompatible with the paramagnetic character of the
quasiparticle Pauli susceptibility, i.e. the susceptibility
enhancement due to impurity bound states is always positive, so
while Friedel-type oscillations can occur, the magnetization is
always aligned with the external field.
Finally, the ``Knight
shift" calculated in Ref. \onlinecite{JChang:2004} is defined to
be a local susceptibility enhancement very near the impurity; in
fact the measured Knight shift in experiments where the nucleus is
distinct from the impurity itself is the shift of the {\it total}
NMR line, determined by sites far from the impurities.

Thus a theoretical calculation which includes  local magnetic
moment formation,  together with quasiparticle impurity bound
states and their interference, is of considerable interest to
understand the simple but striking features of the NMR experiments
in the superconducting state\cite{SOuazi:2006,JBobroff:2001}.
 A complete
theory of these phenomena must be able to account not only for the
Knight shift, but for  the detailed behavior of the site-specific
NMR lines produced by the different nuclei probed in different
experiments.
%
In this paper, we study Zn ions modelled as strong potential
scatterers in a $d$-wave superconductor with AF correlations
treated within a weak coupling approach as in Ref.
\onlinecite{YOhashi:2002}.  In Section \ref{impurity} we describe
 the model for a single impurity in a $d$-wave
superconductor with correlations and the magnetization it induces,
then in Section \ref{manyimp} study interference effects on the
magnetization distribution when many impurities are present.
 In Section
\ref{impNMR} we combine the predicted impurity contribution to the
NMR linewidths with  vortex effects,  and compare to the Ouazi
{\sl et al.}\cite{SOuazi:2006} $^{17}$O NMR experiment.   Section
\ref{Li} is devoted to the application of  the same results to
compare with  experiments using the $^7$Li nucleus. In Section
\ref{conclude}, we present our conclusions and implications for
other experiments on the cuprates, as well as new questions raised
by our interpretation.

\section{Single impurity in system with antiferromagnetic correlations}
\label{impurity}

\subsection{Formalism}
\label{formalism}

A strong nonmagnetic impurity in the presence of antiferromagnetic correlations will
 induce a pattern of local staggered magnetization with maximum peaks on nearest-neighbor sites.
  When many impurities are present, these local states interfere, washing out nearest-neighbor
   peaks and producing a smooth distribution of local fields.  The NMR line shift and broadening
   is caused by this impurity effect in conjunction with the vortex contribution.  In order to
    study this magnetic behavior, we begin with a two-dimensional tight-binding Hamiltonian of
    a {\it d}-wave superconductor with AF correlations treated within mean field theory:
\begin{eqnarray}\label{Hamiltonian}
\hat H = &-&\sum_{ij\sigma}t_{ij}\hat c_{i\sigma}^\dagger\hat c_{j\sigma} + \sum_{i\sigma}
\left( Un_{i-\sigma} + \epsilon_{i\sigma} - \mu\right)\hat c_{i\sigma}^\dagger\hat c_{i\sigma}
\nonumber\\
&+& \sum_{i\delta}\left( \Delta_{\delta i}\hat c_{i\uparrow}^\dagger
\hat c_{i+\delta\downarrow}^\dagger + \mbox{H.c.}\right),
\end{eqnarray}
where the hopping term includes nearest-neighbor hopping $t$ and
next-nearest-neighbor hopping $t'$, $\epsilon_{i\sigma}\equiv
V_{imp}-g\mu_B {1\over 2} B \sigma$ describes  the  impurity and
Zeeman site energies, $\mu$ is the chemical potential,
$\Delta_{\delta i}$ is the nearest neighbor pairing potential,
 and $\delta\in\left\{{\bf \hat x, -\hat x, \hat y, -\hat y}\right\}$ are
 unit lattice vectors to nearest-neighbors.  Here $g\approx 2$ is the electron g-factor, $\mu_B$
    is the Bohr magneton, and $B$ is the applied field along the $c$-axis.  The electron number and {\it d}-wave (singlet)
  pairing parameters are defined as $n_{i\sigma} = \langle \hat c_{i\sigma}^\dagger\hat c_{i\sigma}
  \rangle$ and $\Delta_{\delta i} = V\left\langle \hat c_{i\uparrow}\hat c_{i+\delta\downarrow} +
  \hat c_{i+\delta\uparrow}\hat c_{i\downarrow}\right\rangle/2$.
We note that the Hamiltonian (\ref{Hamiltonian}) has been used
extensively  to study bulk competing phases, disorder and vortex
induced magnetization, as well as novel bound states at interfaces
between antiferromagnets and superconductors\cite{allHamiltonian}.

  Eq. (\ref{Hamiltonian})\,\,
   can  be diagonalized by using the Bogoliubov transformation.  The corresponding Bogoliubov--de
    Gennes equations must be solved iteratively until a self-consistent solution is found:
\begin{eqnarray}
\left(\begin{matrix}\hat\xi_{\uparrow}&\hat\Delta\\\hat\Delta^*&-\hat\xi_{\downarrow}^*\end{matrix}
\right) \left(\begin{matrix}u_{n}\\v_{n}\end{matrix}\right) = E_{n}\left(\begin{matrix}u_{n}\\v_{n}
\end{matrix}\right),
\end{eqnarray}
where positive eigenvalues correspond to spin-up excitations and negative eigenvalues correspond to
spin-down excitations.  The matrix operators are defined by $\hat\xi_{\sigma}u_{n,i} = -\sum_{ij}
t_{ij}u_{n,j} + \left( Un_{i-\sigma}+\epsilon_{i\sigma} - \mu\right)u_{n,i}$ and $\hat\Delta u_{n,i}
= \sum_{\delta}\Delta_{\delta i}u_{n,i+\delta}$.  The mean field parameters, updated after each
iteration until sufficient convergence is achieved, can be computed by:
\begin{eqnarray}
\Delta_{\delta i} = {V\over 4}\sum_{n}\left(u_{n,i}v_{n,i+\delta}^* + u_{n,i+\delta}v_{n,i}^*\right)
\tanh{\left(E_n\over 2k_BT\right)},
\end{eqnarray}
and $n_{i\sigma} = (n_i + \sigma m_i)/2$, where $n_i$ is the average electron density at site $i$
and $m_i$ is the magnetization on site $i$:
\begin{eqnarray}
n_i &=& 1 - {1\over 2}\sum_{n}\left(|u_{n,i}|^2 - |v_{n,i}|^2\right)\tanh{\left(E_n\over 2k_BT\right)}
,\\
m_i &=& -{1\over 2}\sum_{n}\left(|u_{n,i}|^2 +
|v_{n,i}|^2\right)\tanh{\left(E_n\over 2k_BT\right)}.
\end{eqnarray}
It is important to note that in this model the superconducting
pairing is taken to be a phenomenological constant, and is not
affected by the Hubbard repulsion $U$.  In this sense results may
differ from true inhomogeneous spin fluctuation models where both
magnetism and pairing are driven by the same correlations.

When deciding which ordered phase is the ground state, it is
necessary to know the total energy $E$ of the system, defined as
${\langle\hat H\rangle}$.  It is given by:
\begin{eqnarray}
E = K.E. + \sum_i \left[\epsilon_{i\uparrow}{n_i+m_i\over 2}+
\epsilon_{i\downarrow}{n_i-m_i\over 2} -\mu n_i +\right.\nn\\\left.
{U\over 4}\left(n_i^2 - m_i^2\right) -\sum_\delta
{\left|\Delta_{\delta i}\right|^2\over V}\right],
\end{eqnarray}
where ${K.E.}$ is the kinetic energy of the system
\begin{eqnarray}
K.E. = -{1\over 2}\sum_{ijn}t_{ij}\left[ v_{n,i}v_{n,j}^{*} -
u_{n,j}u_{n,i}^{*}\right]\tanh\left( {{E_{n}}\over
{2k_{B}T}}\right).
\end{eqnarray}
We will often speak of the total magnetic moment (or spin) of the
system.  This is given (in units of ${\hbar}$) by
\begin{eqnarray}
S_z = {1\over 2}\sum_{i} m_i.
\end{eqnarray}
We will also make reference to the {\it d}-wave order parameter,
defined as
\begin{eqnarray}
d_i = {1\over
4}\left[\Delta_{x,i}+\Delta_{-x,i}-\Delta_{y,i}-\Delta_{-y,i}\right].
\end{eqnarray}
In the following we operate at a band filling $n= 0.85$ such that
the ground state of the homogeneous system is always pure $d$-wave
of the form $\Delta_\k\propto \cos k_x - \cos k_y$.

Finally, we have occasional need for spatially resolved spectral
information, and so also calculate the LDOS $N(E,i)$, via
\begin{equation}
N(E,i)\! = {1\over 2}\!\sum_{n} \!\! \left(
\left|u_{n,i}\right|^2\delta\left(E - E_{n}\right) +
\left|v_{n,i}\right|^2\delta\left(E + E_{n}\right)\right).
\end{equation}
The chemical potential $\mu$ is adjusted to produce an average
electron density of 0.85, corresponding to 15\% hole doping
characteristic of optimally doped cuprates. We give all energies in
units of $t$ and set $t' = -0.2$,  to mimic typical Fermi surface
shapes found in these systems, and $V = 1$ to give a critical
temperature of $0.15t$.

\begin{figure}[t]
\begin{center}
\leavevmode
\includegraphics[clip=true,width=.99\columnwidth]{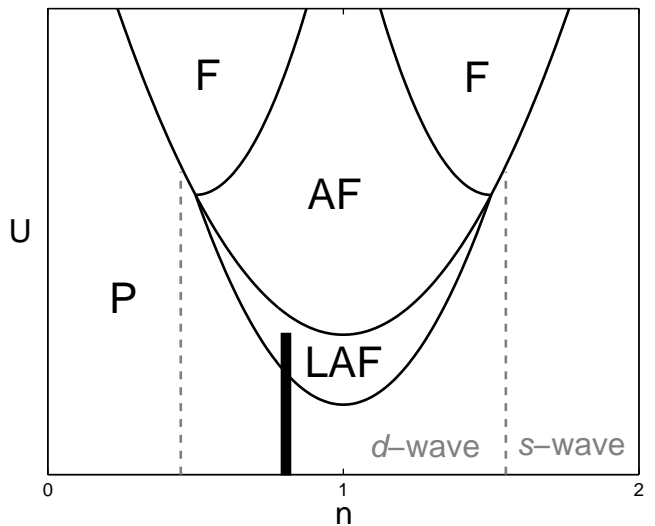}
\caption{$T=0$ $U$-n phase diagram for Hubbard  plus
nearest-neighbor pairing model treated in mean field used in this
work. F, AF and P are the usual ferromagnetic, antiferromagnetic,
and paramagnetic phases found for the homogeneous Hubbard model. LAF
is the local antiferromagnetic phase where spontaneous staggered
moments form around a single potential scatterer. The  gray lines
indicate the boundaries of coexistence of $d$- and $s$-wave
superconductivity with the magnetic phases. Calculations are
performed in the dark rectangle region between $0<U<2.0$ and at 15\%
hole doping, $n=0.85$.} \label{fig:phasediag}
\end{center}
\end{figure}

\begin{figure}[t]
\begin{center}
\leavevmode
\includegraphics[width=0.8\columnwidth]{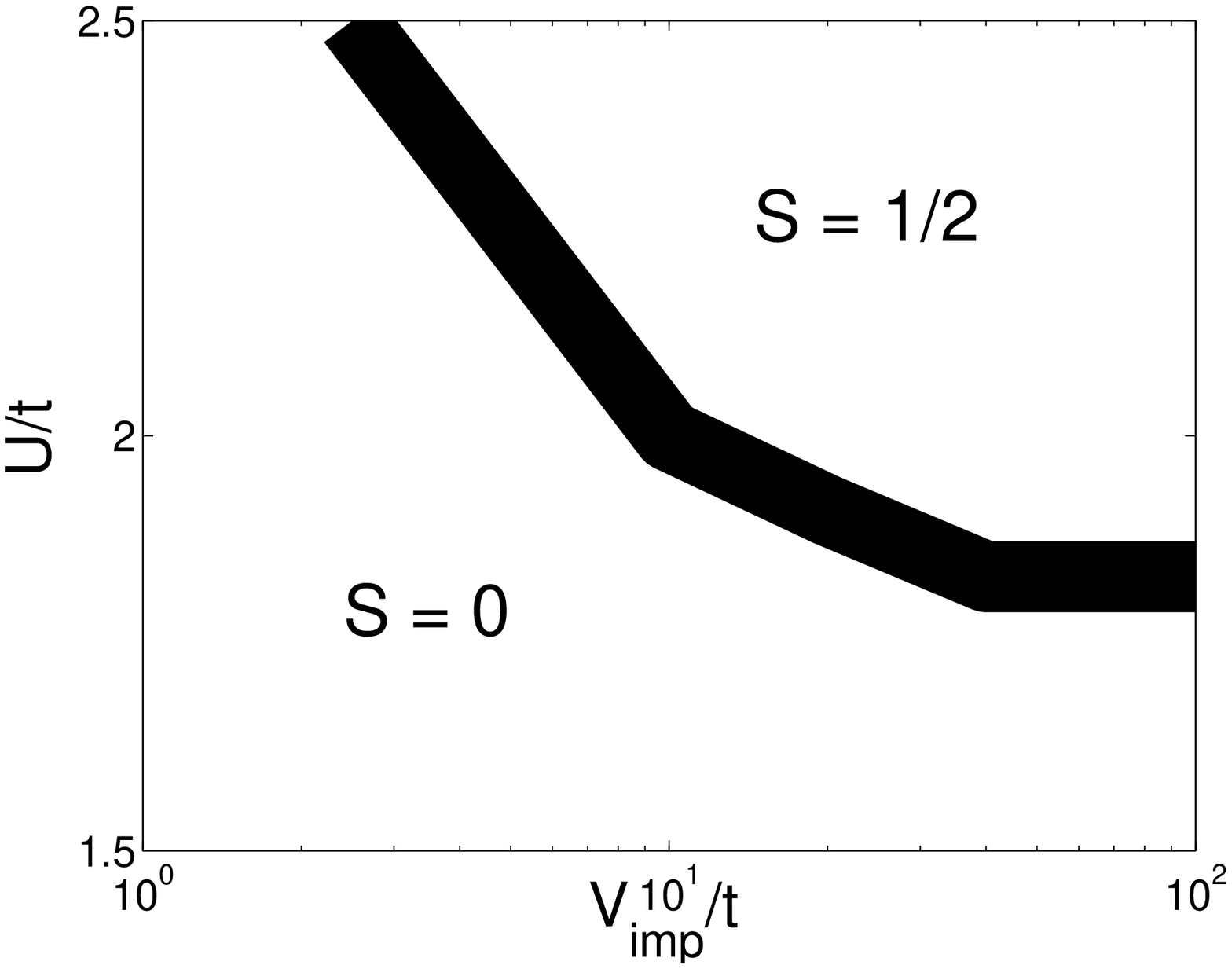}
\includegraphics[clip=true,width=.99\columnwidth]{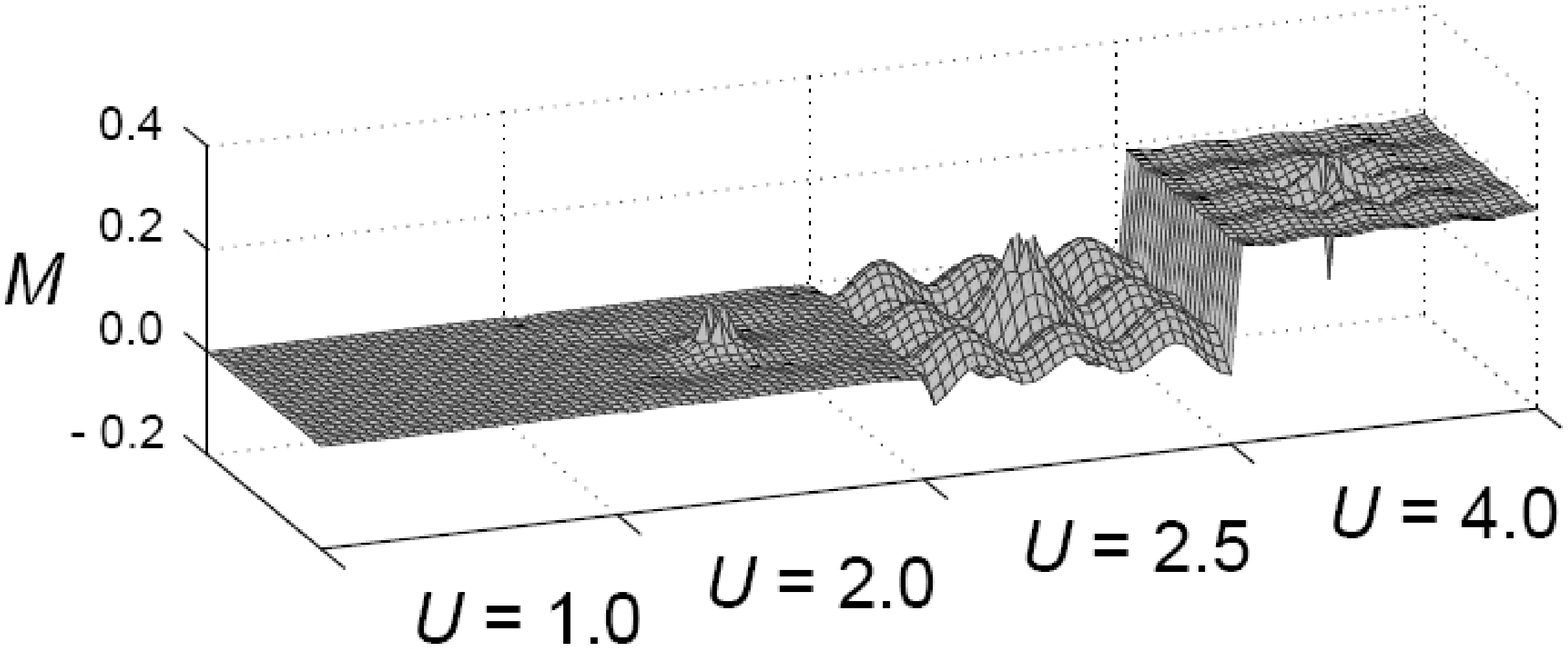}
\caption{Top: phase diagram for a single impurity as a function of
the  impurity potential $V_{imp}$ and $U$, as determined by the
presence of a nonzero magnetization at $T=0.013$.  Note there are
only two phases, $S_z$=0 and $S_z=1/2$. Bottom: For strong
impurity with $V_{imp}=100$, dependence on U of the staggered
magnetization, defined as $M_i = (-1)^i m_i$, for a 28 $\times$ 28
system. The system is completely nonmagnetic ($M$ = 0) for small
values of U, assumes a local staggered state for intermediate
values, and saturates toward a roughly homogeneous AF phase for
large values. For the cases with nonzero magnetization, $S_z =
1/2$. The "wave-like" AF ordering in the bulk for U = 2.5 and 4.0
is due to finite size effects and the periodic boundary conditions
of the system.} \label{fig:localmoment_transition}
\end{center}
\end{figure}

\begin{figure}[t]
\begin{center}
\leavevmode
\includegraphics[clip=true,width=.99\columnwidth]{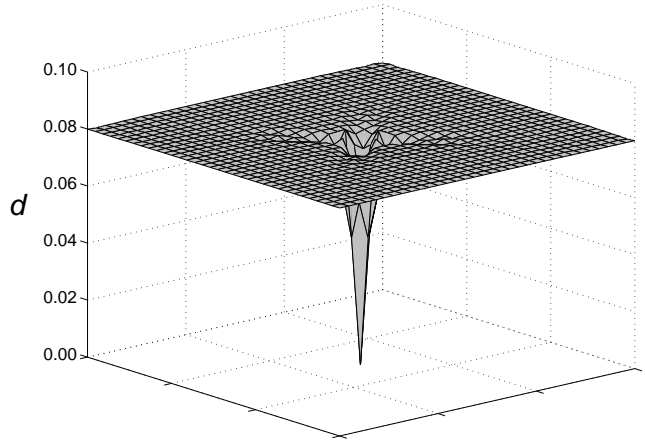}
\caption{$d$-wave order parameter suppression at an impurity of
strength $V_{imp}=100$ for $U=1.75$, $V=1$,  34x34 system, $g \mu_B
B/2$= 0.004, and $T$ =0.013.} \label{fig:d-wave_ops}
\end{center}
\end{figure}

\begin{figure}[t]
\begin{center}
\leavevmode
\includegraphics[clip=true,width=.99\columnwidth]{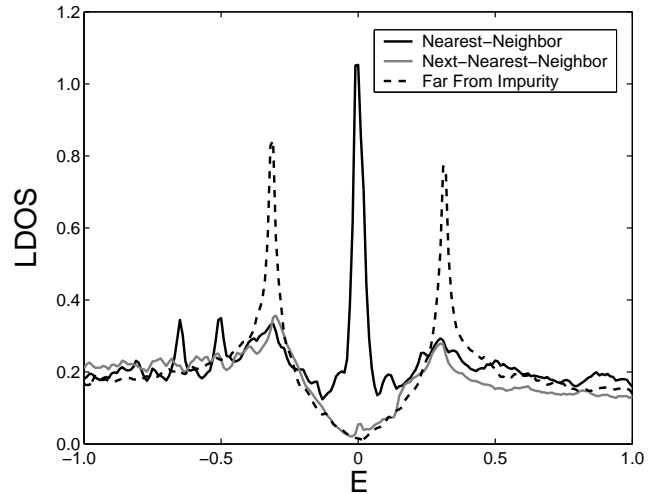}
\caption{The LDOS for a single-impurity system displaying
paramagnetic behavior ($U$ = 1.25,  $T$ = 0.013, $g\mu_B
B/2$=0.004), computed using a 28$\times$ 28 system and 20 $\times$
20 supercell.  This result is in agreement with past studies using
a similar model \cite{YChen:2004}. } \label{fig:LDOS}
\end{center}
\end{figure}

\subsection{Single impurity phase diagram}

\label{1impphasediag} In Fig. \ref{fig:phasediag}, we give a
schematic phase diagram for the model (\ref{Hamiltonian}) with
crude phase boundary lines and details suppressed for simplicity.
Here, ``AF", `F", and ``P" denote self-consistent mean-field
phases of Eq. (\ref{Hamiltonian}) characterized roughly as
antiferromagnetic, ferromagnetic, and paramagnetic, respectively.
For example within the "AF" region it is well-known that spin
density waves with ordering vectors other than exactly
${\mathbf{Q}}=(\pi,\pi)$ can be stabilized with details depending
on the doping level and band-structure parameters. Since we have
included a separate nearest-neighbor pairing interaction term $V$
in the Hamiltonian in order to study the superconducting state, we
have also indicated in Fig. \ref{fig:phasediag} the regions of
doping over which nearest neighbor $d$-wave or $s$-wave pairing
symmetry characterizes the ground state.  Note that in this case
the ordered magnetic phases coexist with superconductivity within
the model; again details have been suppressed for simplicity, and
because we are here primarily concerned with the paramagnetic
($d$-wave superconducting) phase.

Upon addition of a single strong impurity potential to the model, we
find new inhomogeneous ground states present.  Of most interest is a
region of local staggered magnetism surrounding the impurity,
referred to as the ``local antiferromagnetic" (LAF) phase.  In this
phase, the impurity-induced staggered magnetization vanishes at a
large distance from the impurity, and the net spin $S_z$ summed over
the whole system is found to be 1/2\cite{YOhashi:2002}.  At larger
$U$, the impurity still generates a net spin 1/2, but the long-range
ordering dominates, and the staggered magnetization has the maximum
polarization arbitrarily far from the impurity. For a given fixed
band structure and doping $n=0.85$, we have plotted in Fig.
\ref{fig:localmoment_transition} the transition line between the
(total) $S_z=0$ state and the $S_z=1/2$ state for varying $U$ and
impurity potential $V_{imp}$; clearly increasing either $U$ or
$V_{imp}$ tends to favor the local magnetic ``spontaneous moment"
state.  In the lower panel of Fig. \ref{fig:localmoment_transition}
we show the staggered magnetization patterns in each of the states
for increasing $U$. Note that in general we do {\it not} find an
intermediate state with $S=0$ {\it and} impurity-induced magnetism,
in contrast to a recent study of the same Hamiltonian in Ref.
\onlinecite{YChen:2004}, and believe the presence of this state to
be an artifact of the particular size system studied by these
authors. We find generically either a $S_z=1/2$ state with local
staggered magnetic order, or a state with no local magnetization at
all in zero field, referred to as $S_z=0$ in Fig.
\ref{fig:localmoment_transition}.   This is consistent with the
results of Wang and Lee\cite{ZWang:2002}, albeit derived in the weak
coupling limit rather than for the $t-J$ model.

We recover other single-impurity results known for this model which
we state for the sake of completeness.  For example, the $d$-wave
order parameter is strongly suppressed in the vicinity of a strong
potential scatterer, mostly over a length scale of one lattice
spacing, but with a longer-range decay envelope over the coherence
length $\xi_0$ (Fig. \ref{fig:d-wave_ops}).   The impurity also
induces a much smaller $s$-symmetry order parameter(not shown), and
gives rise to a strong resonance in the LDOS near the Fermi level.
The spatial intensity of the resonant state is centered primarily on
the nearest-neighbor sites of the impurity (Fig. \ref{fig:LDOS}), in
apparent conflict with the most naive interpretation of STM
experiments; there are several competing explanations why this could
be so. These phenomena have been reviewed and references given in
\cite{AVBalatsky:2006} (see also \cite{AndreevAndersen}).

\subsection{Response of $S_z=0$ state to applied field}
\label{paramagnetic1imp}

In this paper we focus primarily on the impurity-induced state
which has net spin $S_z=0$ in zero field.  This is because there
is no evidence for impurity-induced magnetization of any kind in
zero field for almost all  of the YBCO phase
diagram\cite{SonierRMP}. NMR experiments, which are sometimes
cited as providing evidence for spontaneous impurity induced
magnetization, are of necessity performed in finite applied field.
Current $B=0$ neutron scattering\cite{Keimernomag} and
$\mu$SR\cite{lackofmoment} measurements find no evidence of
ordered static magnetization at any wavevector.  Furthermore, both
NMR and direct susceptibility measurements indicate that the
induced states are paramagnetic, i.e. the magnetization vanishes
proportional to the applied field. We note that there is
considerable recent evidence that the situation is different in
LSCO, where static magnetism appears to exist even without an
applied external field at low
temperatures\cite{BLake:2002,HKimura:2003}.   This may also
explain unusual transport properties in LSCO compared to
YBCO\cite{AHDresden,Kontani}.

\begin{figure}[t]
\begin{center}
\leavevmode
\includegraphics[clip=true,width=.99\columnwidth]{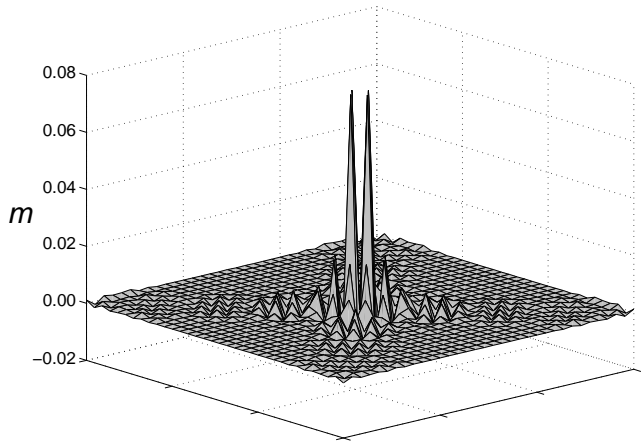}
\caption{The field-induced local magnetization for $U = 1.75$,
$g\mu_B B/2 = 0.004$ and $T = 0.013$ on a 34$\times$34 system. The
total moment of the system is $\langle S_z\rangle = 0.294$. Weak
long-range AF correlations extend out from the impurity along
45$^\circ$ diagonals.  } \label{fig:localmoment}
\end{center}
\end{figure}

With these considerations in mind, we choose a value of $U$ which
will induce significant antiferromagnetic correlations close to
half-filling, but  is not sufficient to cause the formation of
magnetic moments around impurities
 in zero field.  The impurity is taken to have
 an on-site spin-independent potential of strength $V_{imp} = 100$, roughly consistent with STM
 at least as far as the energy of the Zn LDOS resonance is concerned.
 To start, we study the Bogoliubov-de Gennes equations for a single such impurity with systems
  of size $34\times34$ with periodic boundary conditions.  The Zeeman response of the electronic
  spins to the applied field is included in the $\epsilon_{i\sigma}$ term in Eq. (\ref{Hamiltonian}), as discussed above.
     In our approach, we do not include the orbital response of
      the system to the applied vector potential, but rather
      attempt to account for the presence of the vortex state
      phenomenologically (see below).  We wish to compare our results with
     the data from the Ouazi {\sl et al.}\cite{SOuazi:2006} experiment; we therefore take $t = 100$ meV and set
      $|g \mu_B B/2| = 0.004$ ($B \approx 7$ T) and $k_BT = 0.013$ ($T \approx 15$
      K  ). Application of such a field induces, as
      expected\cite{NBulut:2000,NBulut:2001,YOhashi:2001,YOhashi:2002},
      a local staggered magnetization of the Cu spins around the
      impurity site, depicted in Fig. \ref{fig:localmoment}.

\begin{figure}[t]
\begin{center}
\leavevmode
\includegraphics[clip=true,width=.99\columnwidth]{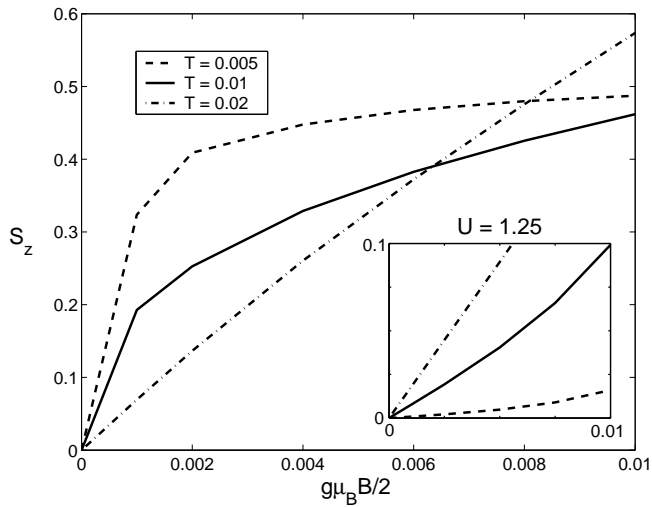}
\caption{The response $S_z$ induced by a field $B$ of a  system with
$U = 1.75$ for three values of $T$. In the limit of zero field, the
system is nonmagnetic regardless of temperature. Inset: field
response  for $U = 1.25$. } \label{fig:fieldresponse}
\end{center}
\end{figure}

      Below we argue that results consistent with experiment
      on optimally doped samples require a value of $U$ close to
      the threshold for the creation of static impurity-induced
      zero-field
      magnetism (i.e. close to the phase boundary in Fig.
      \ref{fig:localmoment_transition}).  This means that the
      field dependence can acquire nonlinearities, as shown in
      Fig. \ref{fig:fieldresponse}.   For most of this work, we
      examine $U=1.75$, although we also exhibit the
      consequences of choosing other values.

The size of $t$ assumed in order to  compare with experiment is
probably a factor of 2-4
    smaller than deduced from ARPES experiments on cuprates, but
    we do not expect this to alter our qualitative conclusions.
   The low value of $t$ is chosen such that reasonably low
   temperatures $T/T_c$ can be accessed without encountering
   finite size effects.  For this choice, together with the choice of pair interaction
   $V$,  $T_c=0.15$ corresponds to
   about 175K, and the  field parameter of  $g\mu B/2 = 0.004$ to the experimental value of 7 T.  The reader should
   therefore take the scales given in Kelvin only as extremely rough
   comparisons.

\begin{figure}[t]
\begin{center}
\leavevmode
\includegraphics[clip=true,width=.99\columnwidth]{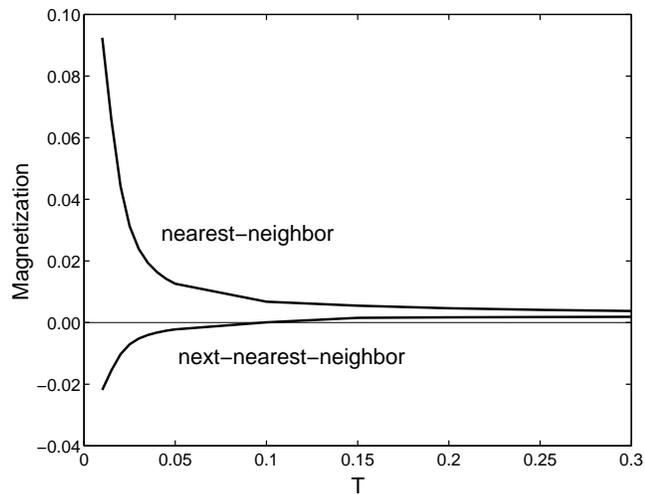}
\caption{Magnetization in fixed field $g\mu_BB/2$=0.004 vs. $T$ on
nearest neighbor  and next nearest neighbor sites for a strong
impurity with $U=1.75$.  } \label{fig:singleimp_S_vs_T}
\end{center}
\end{figure}

In the linear field response regime (Fig.
\ref{fig:fieldresponse}),  our calculations should be very similar
to those of Ohashi for the nearest-neighbor weak-field
susceptibility\cite{Ohashicomment}.   Thus it is not surprising
that we also find a strong increase of the magnetization on the
nearest neighbor sites as the temperature is lowered. This
increase is weak in the normal state, then slows slightly at the
superconducting transition as the gap opens, as depicted in Fig.
\ref{fig:singleimp_S_vs_T}. As the temperature is lowered further,
the resonant state in the $d$-wave superconductor forms (Fig.
\ref{fig:LDOS}), driving the susceptibility to a large value,
which however is expected to saturate at $T\rightarrow 0$, as
indicated in the Figure. We  are unable to calculate results
accurately below a temperature $T\simeq D(a/L)^2$, where $D\approx
8t$ is the bandwidth, because the thermal energy becomes of order
the level spacing in the finite size system.  For the simulations
reported here, the cutoff is of order $T_{min}\sim 0.01t$.  The
next nearest neighbor susceptibility is also enhanced, but has the
opposite sign because the correlations are antiferromagnetic.

\begin{figure}
\begin{center}
\leavevmode
\includegraphics[width= .9\columnwidth]{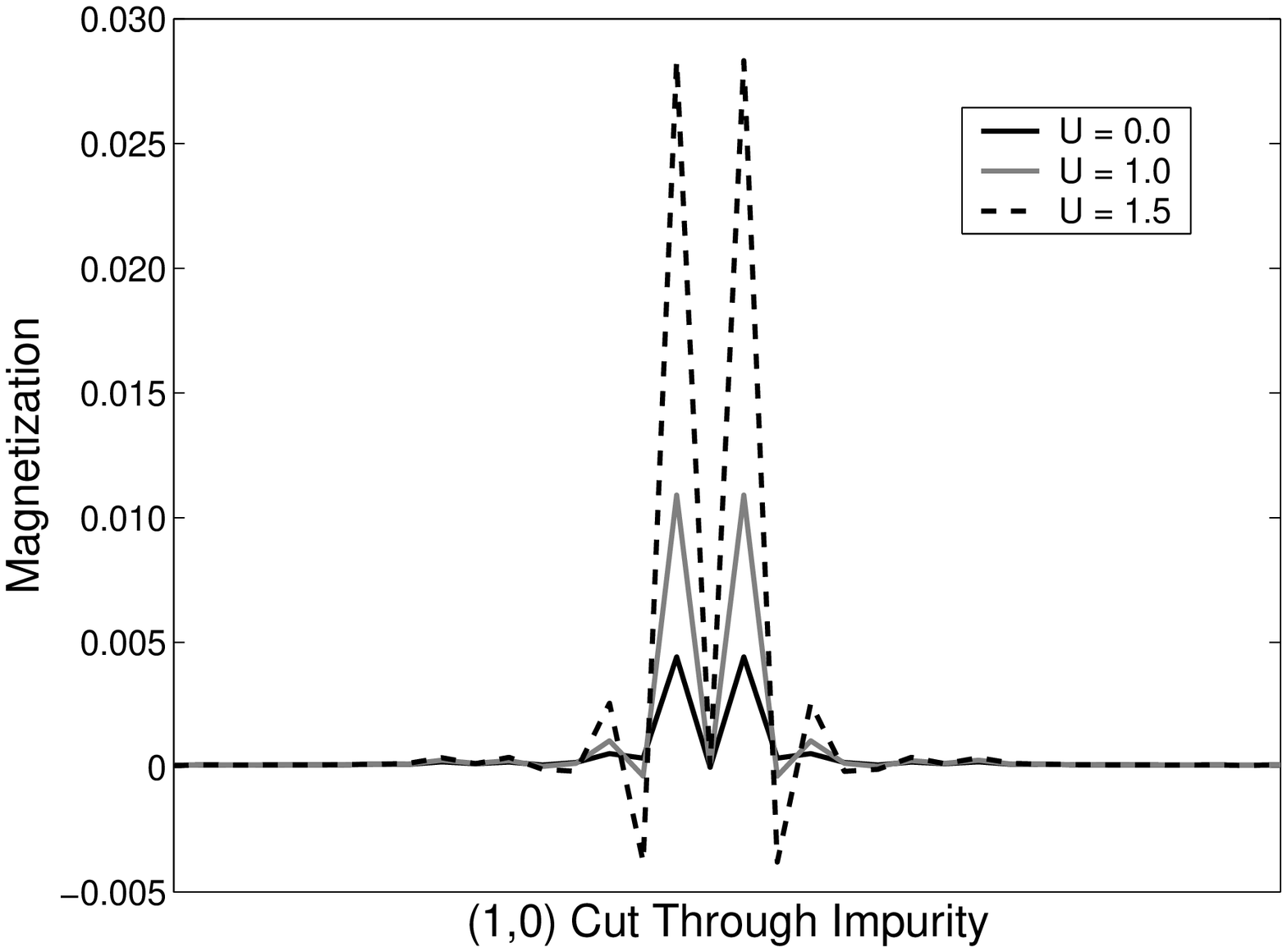}
\includegraphics[width= .9\columnwidth]{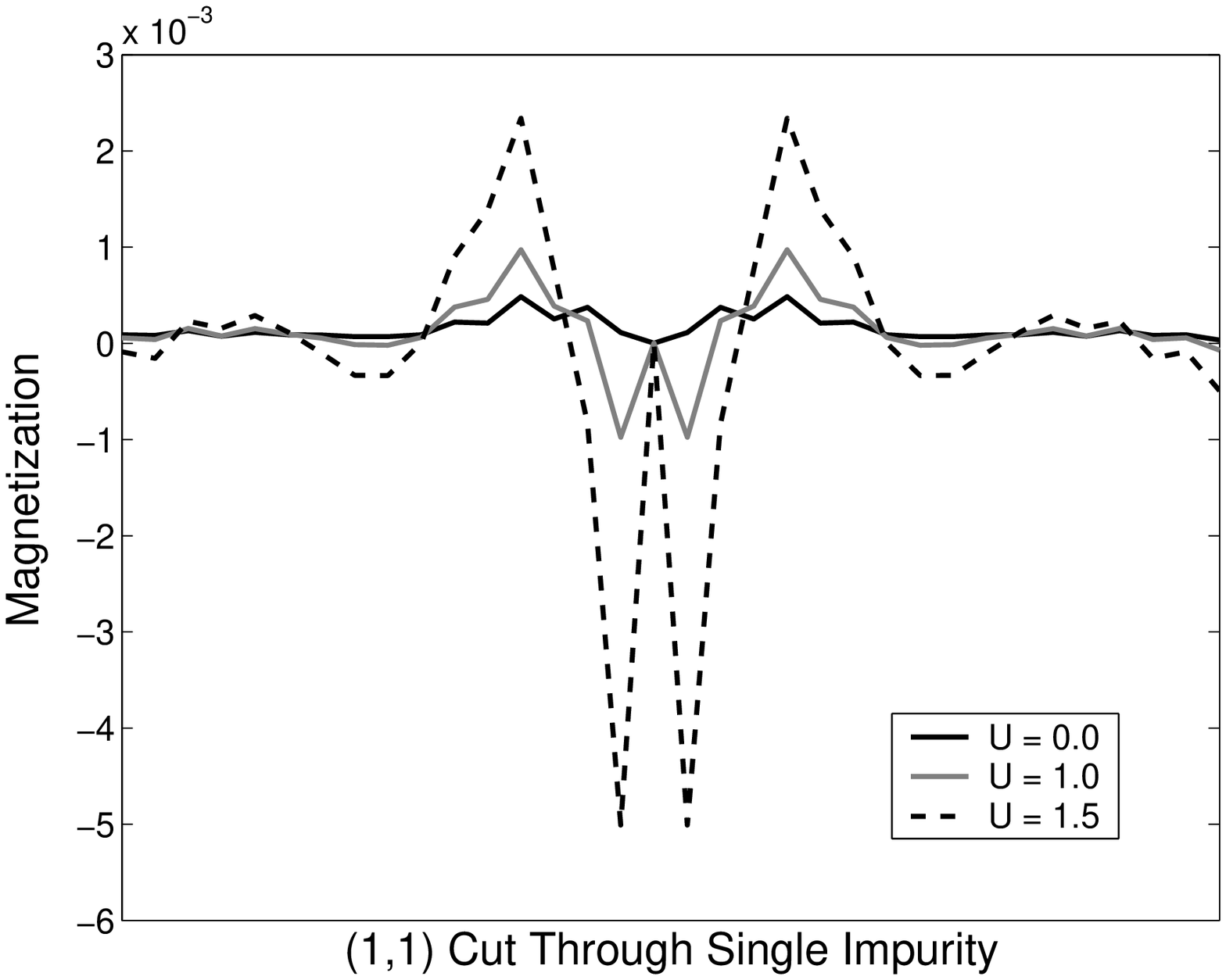}
\caption{ Top: magnetization along (1,0) direction through impurity
with on-site potential $V_{imp}=100$, $g\mu_B B/2=0.004$,
$U=0,1,1.5$, $T=0.013$.  Bottom: same but along the (1,1) direction.
 } \label{fig:mag_mod_U}
\end{center}
\end{figure}

It is worth noting that, in the presence of the nonzero external
field, a modulated local magnetic state is present even in the
absence of the antiferromagnetic correlations driven by $U$.  This
is the analog, in the $d$-wave superconducting state, of
Friedel-like spin density oscillations which represent the
response of the normal metal to a local perturbation.  As such,
the oscillations necessarily take place at an incommensurate wave
vector $2k_F$ which is however close to $(\pi/a,\pi/a)$ because
the system is close to half-filling.  Note that these $U=0$
magnetization oscillations, shown in Fig. \ref{fig:mag_mod_U}, are
driven by the Pauli susceptibility in the $d$-wave superconductor.
For a weak impurity, this  response is quite weak at low
temperatures, due to the linear $\omega-$dependence of the
$d$-wave density of states near the Fermi level.  On the other
hand, the LDOS resonance at the Fermi level in the case of a
strong impurity enhances this local response substantially.  Note
that the magnetization is always positive, however, since the
local susceptibility is proportional to an enhanced LDOS at the
Fermi level.  When correlations are added, as indicated by the
increasing $U$ in Fig. \ref{fig:mag_mod_U}, the response can be
many times that of the pure BCS system with noninteracting
quasiparticles, and takes on an alternating character, as seen.
These effects will lead to asymmetries in NMR lineshapes, as
discussed below.  We note further that the ``background"
homogeneous magnetization of the system in nonzero external field
is present in  Fig. \ref{fig:mag_mod_U}, but barely visible due to
the small value of the homogeneous $d$-wave susceptibility at low
$T$.

\section{Many-impurity magnetization}
\label{manyimp}

\begin{figure}[t]
\begin{center}
\leavevmode
\includegraphics[clip=true,width=.99\columnwidth]{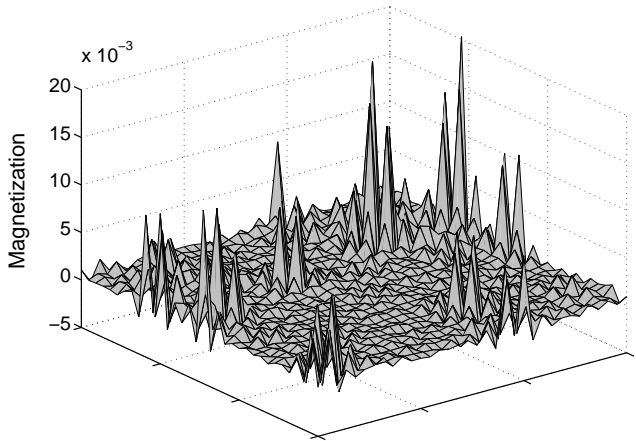}
\caption{Magnetization $m$ for one configuration of a system with
1.5\% random impurities and $U$ = 1.25, $T = 0.013$, $B = 0.004$.}
\label{fig:many_imp_mag}
\end{center}
\end{figure}

In the presence of many strong impurities, the wave functions of
 electrons bound to the impurity interfere at long distances,
leading to collective behavior which is no longer describable by the
1-impurity model.  These effects have been studied in $d$-wave
superconductors without antiferromagnetic
correlations\cite{AVBalatsky:1996,UMicheluchi:2002,DKMorr:2002,LZhu:2003,BMAndersen:2003,WAAtkinson:2003,BMAndersenalone:2003}
The interference of the many impurity states leads to a splitting of
bound state energies and an accumulation of low-energy
impurity-induced energy eigenvalues which are spread out over a
so-called ``impurity band".  In the $d$-wave case, the formation of
the impurity band and the corresponding quasiparticle localization
problem are strongly influenced by the fact that significant
overlaps between two impurity states can take place only if the
impurities are ``oriented" with respect to one another such that the
nodal quasiparticle wavefunctions overlap along the 110 direction.
In fact, analysis of the two-impurity problem show that interference
effects can take place over many tens of lattice spacings  between
optimally oriented
impurities\cite{LZhu:2003,DKMorr:2002,BMAndersen:2003}.  In systems
with per cent level disorder, however, these effects are significant
also for pairs of impurities aligned along the 100 direction.

\begin{figure}[t]
\begin{center}
\leavevmode
\includegraphics[clip=true,width=.75\columnwidth]{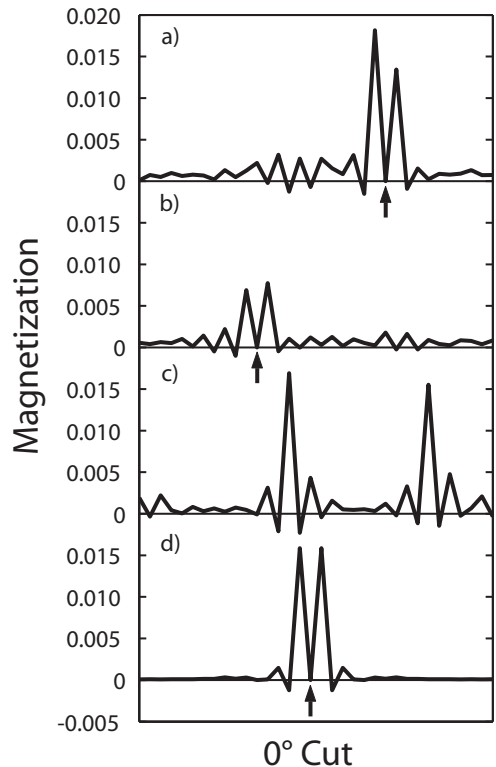}
\caption{0$^\circ$ magnetization cuts through  1\% impurity
systems (a-c)and a single isolated  impurity (d),   taking $U =
1.25$, $T = 0.013$, $g\mu_BB/2 = 0.004 $. Arrows indicate a cut
passes through an impurity site, where magnetization is zero, and
large peaks occur on nearest-neighbor sites. Interference effects
can both enhance and suppress nearest-neighbor magnetizations
relative to the 1-impurity result shown at bottom. Note the cut in
panel (c) does not pass through an impurity site, but passes
through two nearest-neighbor sites.} \label{fig:interference}
\end{center}
\end{figure}

 In the
presence of correlations, interference effects extend to the
magnetic channel and are enhanced by increasing $U$. In Fig.
\ref{fig:many_imp_mag} we show a  system in applied field in the
presence of many strong impurities.  It is clear that the size of
the magnetization on nearest neighbor sites varies significantly
according to the local disorder environment.  To clarify this, we
compare a few of these impurities in the many-impurity sample with
the comparable impurity in isolation in Fig.
\ref{fig:interference}.

\begin{figure}[t]
\begin{center}
\leavevmode
\includegraphics[clip=true,width=.99\columnwidth]{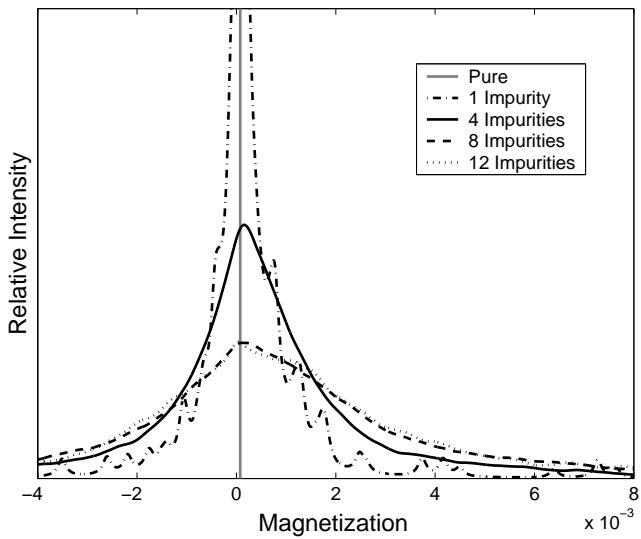}
\caption{Magnetization histogram for a 34 $\times$ 34 system with
$U$=1.75, and 1,4,8 and 12 impurities at $T=0.013$ and $g\mu_B
B/2=0.004$.}. \label{fig:impMags}
\end{center}
\end{figure}

The distribution of magnetizations shown in Fig.
\ref{fig:many_imp_mag} represents all the information necessary to
calculate the NMR response within our approach, since it
determines the distribution of nuclear spin precession
frequencies, in different linear combinations depending on the
locations of the nuclei relative to the Cu sites.  We focus in
more detail on the actual NMR lines in Sections \ref{impNMR} and
\ref{Li} below.  For now, we are interested in showing how
interference effects influence the ``bare" distribution of
magnetizations.  To this end, we collect all the magnetizations in
the system in a histogram for a single impurity, then compare to
progressively larger impurity concentrations in Fig.
\ref{fig:impMags}.  For a single impurity, small satellite peaks
are visible in the spectrum since the same magnetization value
appears on all  sites with fourfold symmetry.  Note that the
satellite peaks associated with the nearest neighbor sites occur
at magnetization values well outside the range of the plot! With
the addition of a few random impurities, these magnetization
values are split, the distribution is smeared, and satellites are
seen to disappear already at sub-percent level concentrations.  In
addition, positive magnetizations are seen to be preferentially
enhanced, due to the density of states effects discussed above.

\begin{figure}[t]
\begin{center}
\leavevmode
\includegraphics[clip=true,width=.99\columnwidth]
{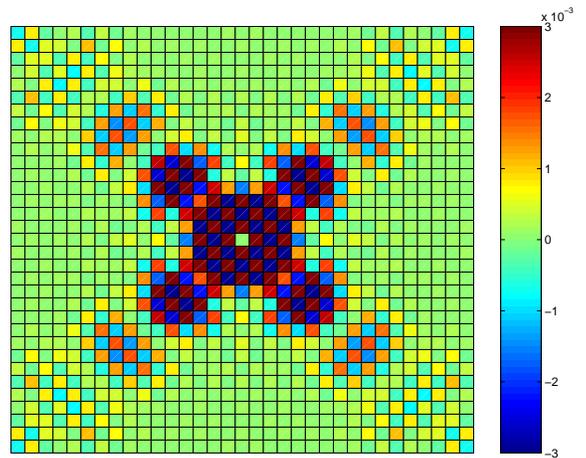} \caption{(Color online) Distribution in
real space of magnetization values around a single $V_{imp}=100$
impurity with $U$=1.75, $T=0.013$, and $g\mu_B B/2=0.004$.}.
\label{fig:impMagsrealspace}
\end{center}
\end{figure}

It is interesting to ask why interference effects in the magnetic
channel are so important in the superconducting state that
satellite features are immediately eliminated.  To this end we
plot the distribution of magnetization values in real space in
Fig. \ref{fig:impMagsrealspace}.  It is seen that those values
which contribute to the satellites close to the peak (which
eventually determine the width of the line for a finite density of
scatterers) are primarily located in the 45$^\circ$ tails of the
quasiparticle wavefunctions some 10-15 lattice spacings from the
impurity site (see below, however).  The orbital parts of these
wave functions are known to interfere strongly in the absence of
correlations provided other impurities are appropriately oriented,
so it is no great surprise that the magnetic parts of these
wavefunctions also strongly interfere.

\begin{figure}[t]
\begin{center}
\leavevmode
\includegraphics[clip=true,width=.99\columnwidth]{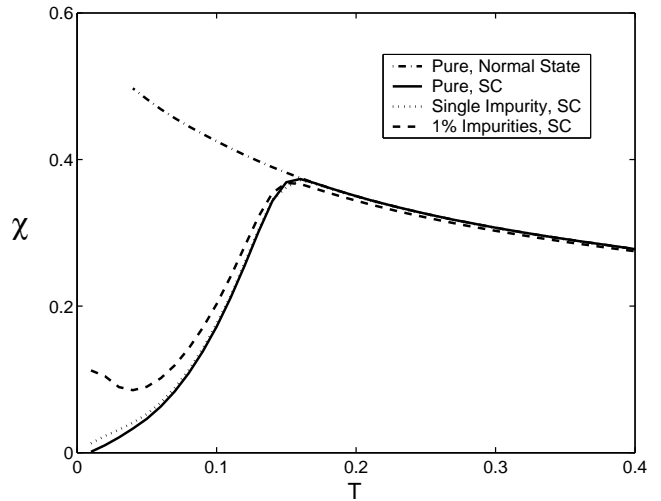}
\caption{Magnetic susceptibility  per site $(1/L^2)dS_z/dB$ for a
28 $\times$ 28 system at $U = 1.25t$ for normal state ($d=0$),
superconducting state, one impurity and 1\% impurities.  }
\label{fig:chi_vs_T}
\end{center}
\end{figure}

Any bulk measurement of magnetization results in an average over
this smeared magnetization distribution.  In addition, the
temperature dependence of the magnetization depend on their position
relative to the impurity.  Contributions from sites far from
impurities decrease with decreasing temperature, as for the
homogeneous $d$-wave superconductor.  Impurity nearest-neighbor
susceptibilities are strongly enhanced, on the other hand.  These
effects combine to determine the total temperature dependence of
thermodynamic properties.  For example, if one measures the total
susceptibility of the sample, it exhibits an upturn at low $T$ if
the  density of impurities is a significant fraction of the sample
(Fig. \ref{fig:chi_vs_T}).

\section{Impurity contribution to $^{17}$O line}
\label{impNMR}

\subsection{Disorder dependence of $^{17}$O line}

The planar $^{17}$O nucleus is situated halfway between two Cu
sites, or between a Cu and a Zn.  It is assumed that it senses the
local field proportional to the sum of the   magnetizations on the
two sites closest to it.  For example,  O nuclei far from the
impurities are subjected to 2$\chi_{hom}B$, where $\chi_{hom}$ is
the susceptibility of the homogeneous system, whereas an O nucleus
next to a Zn atom is subjected to $(\chi_{nn}+0)B$, since there is
effectively zero electron density on the Zn site by assumption.
Nuclei at varying distance from the impurity will measure
different combinations of local magnetizations.

In an NMR experiment, the resonance frequency $\nu $ of a nucleus
is given by:

\begin{equation}
\nu =\frac{\gamma }{2\pi }B\left[ 1+K_{orb}+K_{spin}\right],
\label{nmrfreq}
\end{equation}
where $\gamma $ is the gyromagnetic ratio of the nucleus, $B$ is
the applied magnetic field, $K_{orb}$ is the T-independent orbital
contribution of valence and inner-shell electrons, and $K_{spin}$
is the spin contribution
from the electrons.\ In a simple metal or in a cuprate above T$_{c}$, $%
K_{spin}$ is proportional to the uniform electronic spin susceptibility $%
\chi _{spin}$ through:
\begin{equation}
K_{spin}=\frac{A_{hf}\chi _{spin}}{\mu _{B}} \label{Knightshift},
\end{equation}
where $A_{hf}$ is the hyperfine coupling between the nucleus and
the
electrons, and $\mu_B$ the Bohr magneton.\ In the specific case of $\ ^{17}$O NMR for planar oxygens, a $%
^{17}$O\ nucleus at position $(x;y)$ is coupled to the spin
susceptibility through its two neighboring Cu, leading
to\cite{Takigawa:1989} :
\begin{equation}
^{17}K_{spin}(x,y)=\frac{^{17}A_{hf}\left[ \chi _{spin}(x+\frac{1}{2}%
;y)+\chi _{spin}(x-\frac{1}{2};y)\right] }{\mu _{B}},
\label{localKnight}
\end{equation}
where we assumed here an oxygen lying along the $x$ axis, with $x$
and $y$ the Cu coordinates in units of the Cu lattice spacing $a$.

\begin{figure}[t]
\begin{center}
\leavevmode
\includegraphics[clip=true,width=.99\columnwidth]{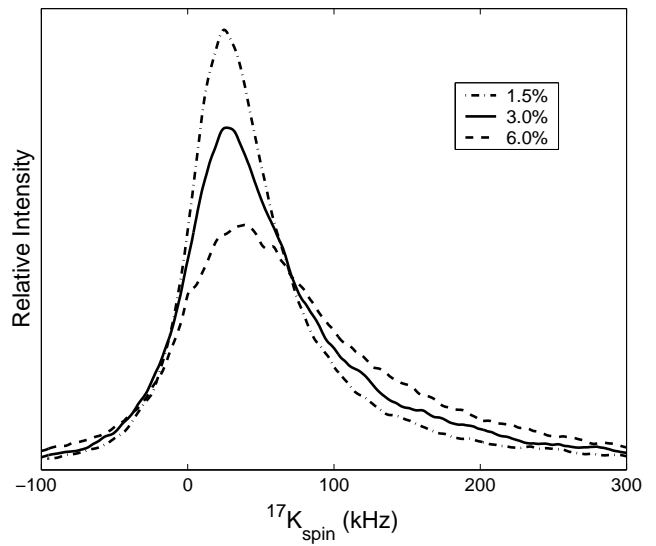}
 \caption{$^{17}$O NMR shift calculated
from Eq. (\ref{localKnight}) for $U=1.75$ for a 34 $\times$ 34
system averaged over 15 random impurity configurations at $g\mu_B
B/2=0.004$ and $T=0.013$.  Numerical smoothing was performed by
convolving the distribution with a Lorentzian of width 3.5kHz.}
\label{fig:NMRshift-1.5,3,6}
\end{center}
\end{figure}

In the pure metallic state above $T_{c}$, $\chi _{spin}$ is
uniform, so
that the NMR consists of a single line shifted by $%
^{17}K_{spin}=2\,^{17}A_{hf}\chi _{spin}/\mu _{B}$. In the
superconducting state, the situation is more complex because of the
presence of the vortex lattice. Eq. (\ref{localKnight}) stays valid,
but $B({\bf r})$ is no longer uniform. As the NMR spectrum is a
histogram of all the frequencies $\nu ({\bf r})$, this spectrum
directly reflects the vortex field distribution, assuming that
orbital and spin shifts stay uniform (see for example Ref.
\onlinecite{APReyes:1997}).

Let us now consider how the impurities affect the NMR spectrum. If
an impurity induces a spatially dependent magnetization as computed
above, it can be regarded as a distribution of the susceptibility
$\chi _{spin}(x;y)$ among Cu sites, leading to a distribution of
$K_{spin}$, i.e.\ a broadening of the NMR\ spectrum. When both
impurities and superconductivity are now taken into account, the NMR
spectrum should  be distributed simultaneously by a distribution of
the local field $B({\bf r})$ due to the vortex lattice and by a
distribution of spin shifts $K_{spin}({\bf r})$ due to the
impurities.\ However, these two distributions are uncorrelated,  as
argued in Ref. \onlinecite{SOuazi:2006}. They should then simply
convolve with each other.

In Fig. \ref{fig:NMRshift-1.5,3,6}, we show how the distribution
of magnetizations is transformed into a distribution of $^{17}$O
frequency shifts due to disorder alone.
 As expected,
increasing disorder broadens the line.  In addition, however, it
is seen that the line asymmetry increases, with enhanced weight on
the positive side.  This effect was indeed observed in experiment,
and attributed to the enhanced spin susceptibility, i.e. density
of states near the Fermi level due to impurities.  Thus some small
line asymmetry would result simply because the density of states
of a disordered $d$-wave superconductor is enhanced in the
interstitial regions far from the impurity, but as we have seen
the susceptibility is also enhanced due to correlation effects,
which are also affected by interference of many impurities.

\begin{figure}[t]
\begin{center}
\leavevmode
\includegraphics[clip=true,width=.99\columnwidth]{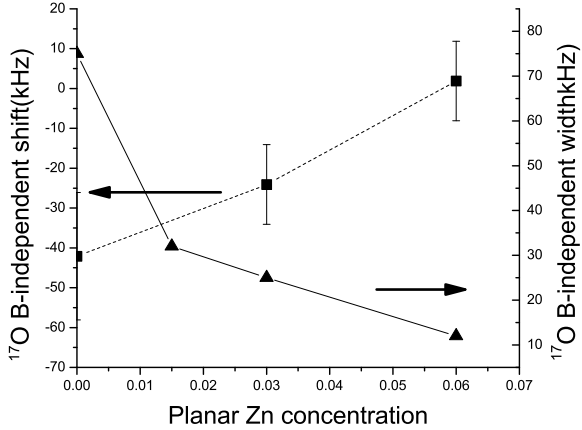}
\caption{Field-independent experimental $^{17}$O shifts (squares)
and widths (triangles) from Ref. \onlinecite{SOuazi:2006}.  Note
no field-independent shift was determined for the 1.5\% sample.}
\label{fig:Ouazi_width_shifts}
\end{center}
\end{figure}

\begin{figure}[t]
\begin{center}
\leavevmode
\includegraphics[clip=true,width=.99\columnwidth]{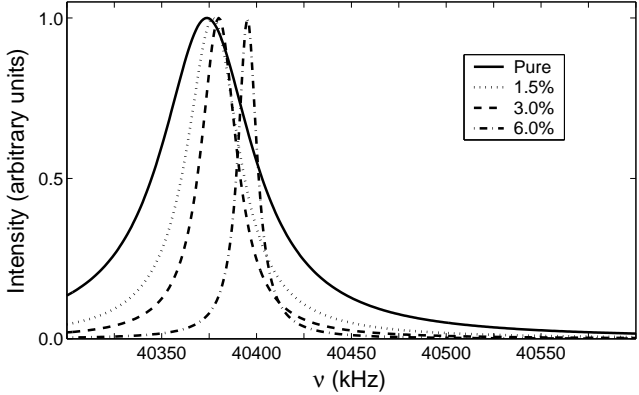}
\caption{Lorentzian vortex field distributions deduced from
field-independent data of Ouazi {\sl et al.}\cite{SOuazi:2006}.
Note the 1.5\% curve was obtained by interpolating the shift of
the 1.5\% sample.  } \label{fig:Lorentzians}
\end{center}
\end{figure}

In order to compare our computation to the experimental NMR lines in
the presence of Zn below $T_{c}$, we must now estimate the effects
of the vortex lattice.   The idea is to produce a series of NMR
lines corresponding to the pure system in the presence of the vortex
lattice field modulation, renormalized by the disorder-enhanced
magnetic penetration depth for each impurity concentration.  These
lines include in principle the effects of spatially varying
superflow, but not of spatially varying spin magnetization.

Simulating the vortex field distribution including quasiparticle
contributions with high accuracy is probably difficult even in the
nominally pure case due to uncertainties regarding the origin and
statistical nature of the disorder in the vortex lattice.
 Following Ref. \onlinecite{SOuazi:2006}, we therefore  identify empirically the vortex-induced part
of the field distribution in the superconducting state with the
field-independent part of the overall NMR shift.  We accept the
determination of the width and shift of the  field-independent
part of the distribution determined by Ouazi {\sl et
al.}\cite{SOuazi:2006}, shown in Fig.
\ref{fig:Ouazi_width_shifts}, and  estimate the vortex
contribution by the Lorentzians determined using these parameters.
The lines thus obtained are shown in Fig. \ref{fig:Lorentzians}.
Note that the data for the 1.5\% Zn sample was not good enough to
extract the field-independent part; we have therefore simply
interpolated linearly between the pure and 3\% sample.

\begin{figure}[t]
\begin{center}
\leavevmode
\includegraphics[clip=true,width=.94\columnwidth]{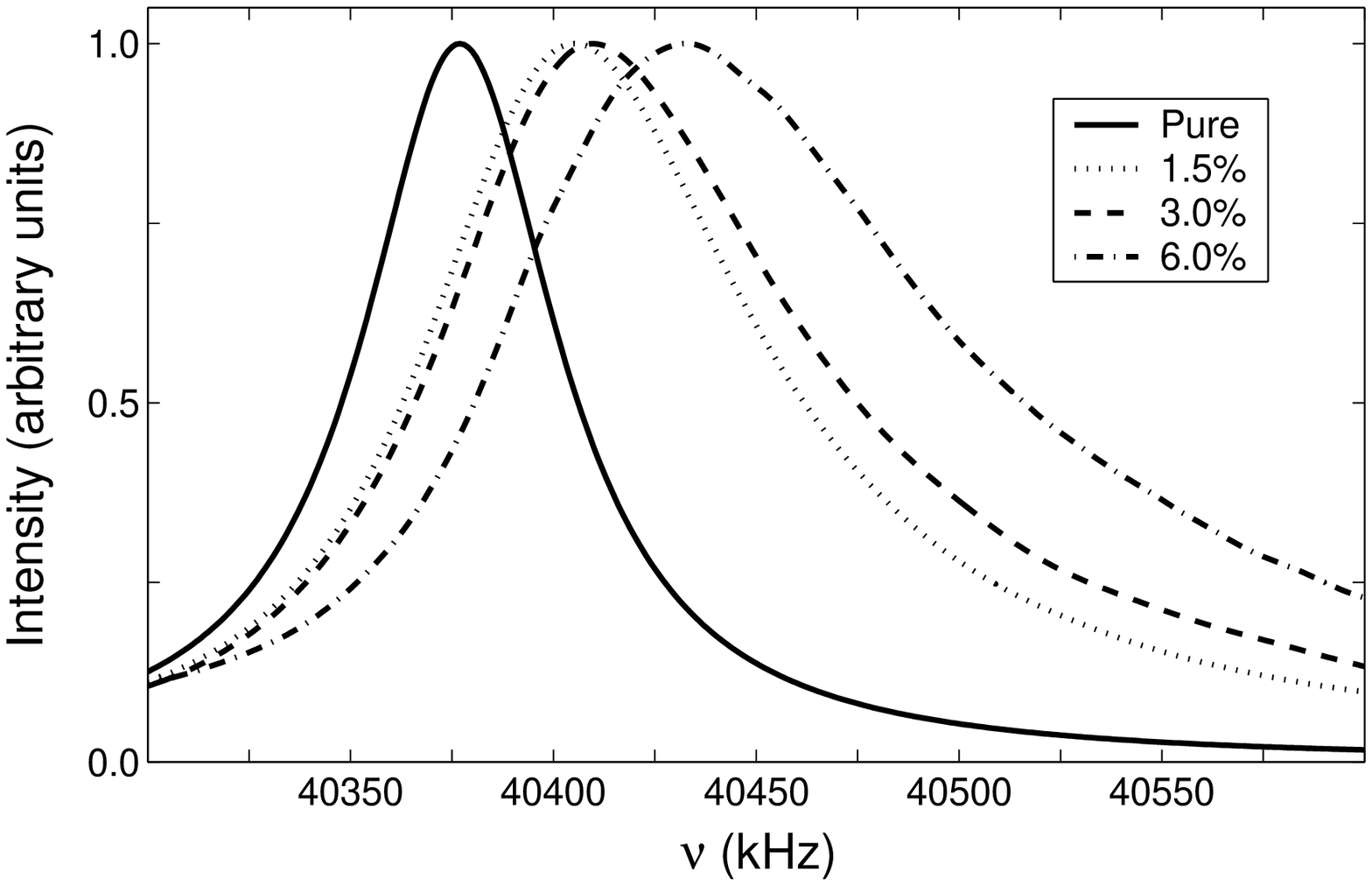}
\includegraphics[clip=true,width=.99\columnwidth]{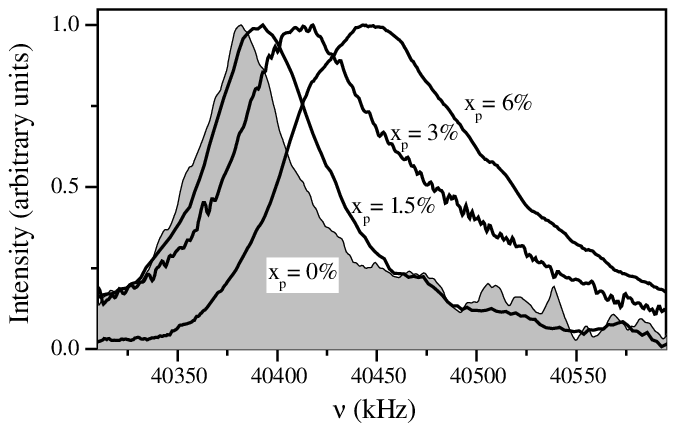}
\caption{Top: normalized theoretical $^{17}$O NMR lines at
$T=0.013$ and $g\mu_B B/2=0.004$ obtained by the procedure
described in the text, for $U=1.75$ and planar Zn concentrations
of $x_p=0\%, 1.5\%, 3\%$, and $6\%$. Bottom: Experimental results
on YBCO powders from Ref. \onlinecite{SOuazi:2006} for the same
concentrations in a 7T field at 15K.} \label{fig:Ouazi_Zn_lines}
\end{center}
\end{figure}

\begin{figure}[t]
\begin{center}
\leavevmode
\includegraphics[clip=true,width=.99\columnwidth]{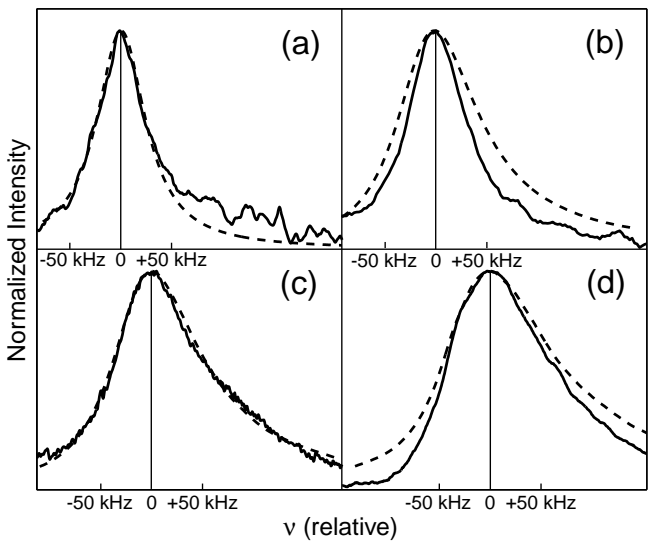}
\caption{ Normalized experimental $^{17}$O NMR lines of Ref.
\onlinecite{SOuazi:2006} at 15K and 7T with all shifts removed
(solid lines) and normalized theoretical impurity NMR lines for
$U=1.75$ convolved with vortex field distributions as described in
text, shifts also removed (dashed lines): (a) pure (b) $x_p =
1.5\%$, (c) $x_p = 3.0\%$, and (d) $x_p = 6.0\%$. }
\label{fig:NMRwidthComparisons}
\end{center}
\end{figure}

These lines are now convolved with the distributions of $^{17}$O
NMR shifts obtained from the impurity effects alone.  This is
justified because in fields of a few Tesla the inter-vortex
distance is of the order of hundreds of \AA, whereas the typical
inter-impurity distance for samples with per cent level Zn is tens
of \AA.  Thus, there can be no significant correlation between the
positions of most of the Zn atoms and the vortices themselves.

 To obtain the corresponding $^{17}$O shifts, we now use
the values $\gamma /2\pi =5.772$ MHz/T, $^{17}A_{hf}=$36kOe
\cite{Yoshinari:1990,Butaud:1990} and $^{17}K_{orb}=0.02\%$
\cite{Takigawa:1989}, which apply for the external field along the
c crystallographic axis as in Ref. \onlinecite{SOuazi:2006}.
Finally we convolve both the impurity and vortex field
distributions. The lines thus obtained are plotted in Fig.
\ref{fig:Ouazi_Zn_lines} and compared to the experimental results
of Ouazi {\sl et al.}\cite{SOuazi:2006} taken at 15K.  The
semi-quantitative variation of the width and asymmetry of the
experimental lines with impurity concentration are seen to be
remarkably well reproduced by the theoretical results. In
addition, the magnitudes of the shifts for different Zn
concentrations are quite well reproduced, with the exception of
the 1.5\% sample, where the shift of the vortex field distribution
was effectively unknown. We note that no extensive fitting in
parameter space was done, so it is quite striking that the
magnitudes and dependence on concentration agree so
quantitatively.

To make further quantitative  comparisons with the widths, which
are the more experimentally reliable quantities, we plot in Fig.
\ref{fig:NMRwidthComparisons} the same normalized lines shown in
Fig. \ref{fig:Ouazi_Zn_lines}, but with the shifts removed.  It is
seen that the theory tracks the increase in width as well as the
overall lineshape extremely well.
\subsection{$T$ dependence of line asymmetry}
\label{lineasym}

Ouazi {\sl et al.}\cite{SOuazi:2006} observed an increase in the
NMR line {\it asymmetry}, with a shift in weight towards the
positive side, as Zn concentration was increased and/or as the
temperature was lowered.  They proposed that this phenomenon was
associated with the formation of the resonant state around Zn
observed by STM, increasing the LDOS at the Fermi level near the
impurity, and thereby enhancing the spin susceptibility.  Since
the LDOS enhancement is always positive, this effect selectively
enhances the broadening on the positive side of the line, provided
it exceeds in magnitude the homogeneous  magnetization in the
regions of the sample far from impurities.  In a $d$-wave
superconductor, this is always true at sufficiently low $T$ since
the bare susceptibility of the clean system vanishes as $\sim T$
at low temperatures. Note that this LDOS-based asymmetry
enhancement is essentially the same effect to which the
``LDOS-only" approaches mentioned above ascribe the entire
enhancement of the local susceptibility measured in NMR. We have
already argued that this is a small effect with regard to the
overall $T$ dependence in the near field of the impurity; here we
show that it can nevertheless play an important role in the
$T$-dependent structure of the line, and in particular the
asymmetry of the linewidth, which arises from contributions
further from the impurities.

We find that the explanation given by Ouazi {\sl et
al.}\cite{SOuazi:2006} is essentially correct, but is strongly
enhanced both by interference between multiple impurities and by
antiferromagnetic correlations. We have seen  that the
susceptibility on the sites nearest the impurity is strongly
increased as the temperature is lowered (Fig.
\ref{fig:singleimp_S_vs_T}), and that this effect is magnified by
increasing $U$ (Fig. \ref{fig:mag_mod_U}). The enhancement of the
asymmetry of the lineshape due to the  interference effect for
fixed temperature was mentioned above, and  illustrated in Fig.
\ref{fig:NMRshift-1.5,3,6}. We therefore anticipate that the
low-$T$ upturns in the asymmetry of the lineshape will be a
characteristic of the present model as well.

\begin{figure}[t]
\begin{center}
\leavevmode
\includegraphics[clip=true,width=.8\columnwidth]{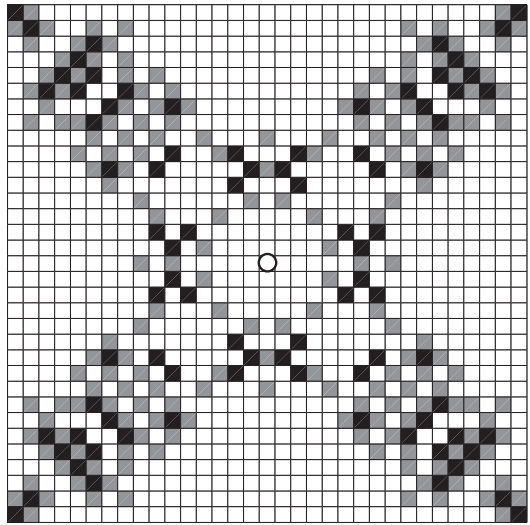}
\caption{Single impurity of strength $V_{imp}=100$ indicated by
circle at center of 34$\times$ 34 system.    Sites with
magnetization values within ranges such as to determine NMR
linewidth for $U=1.75$, $T=0.01$, $g\mu_B B/2$=0.004. Sites with
$-0.0015\le m \le -0.0001$ (``$\nu_L$ sites") are colored black,
and those with $0.0005\le m\le 0.002$ (``$\nu_H$ sites") are
colored gray.
 } \label{fig:key_sites}
\end{center}
\end{figure}

To see the origin of the asymmetry enhancement at low $T$, let us
first examine the temperature dependence of the magnetization of
those sites which actually determine the measured width.  This is
not a completely straightforward proposition, given that the
magnetization patterns in the $d$-wave superconducting state are
determined by a combination of normal Friedel oscillations, local
magnetic correlations, $d$-wave pair correlations, and
interference in the many-impurity case.  Even for a single
impurity, the first three effects combine to make it difficult to
specify, e.g. a given distance from the impurity which is
important for determining the linewidth.  It is clear that it is
not the nearest- or next-nearest- neighbors which do so, but as
shown in Fig. \ref{fig:key_sites}, the set of sites actually
contributing to the positive and negative half-widths -- while
indeed clustered around a range 10-15 lattice spacings from the
impurity in the nodal tails of the wavefunctions, as noted above
-- form a more complicated pattern.

\begin{figure}[t]
\begin{center}
\leavevmode
\includegraphics[clip=true,width=.8\columnwidth]{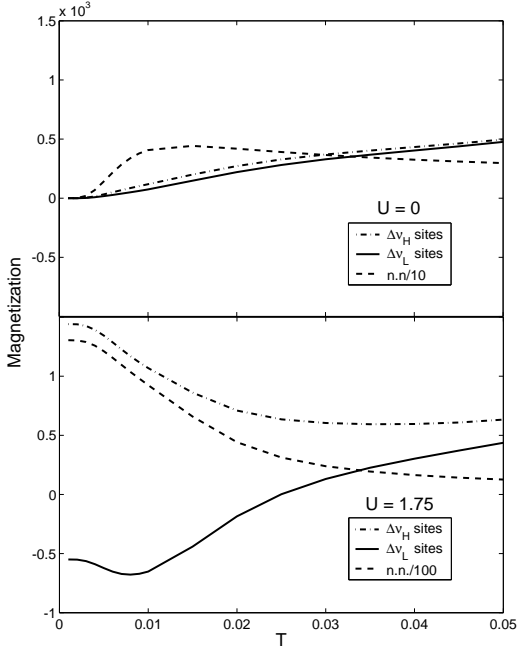}
\caption{Temperature dependence of magnetization on sites
identified as contributing to determination of NMR linewidth, as
specified in Fig. \ref{fig:key_sites}.   }
\label{fig:Tdep_key_sites}
\end{center}
\end{figure}

The temperature dependence of these selected sites is now shown in
Fig. \ref{fig:Tdep_key_sites}, where the influence of electronic
correlations is also illustrated by comparing $U=1.75$ and $U=0$.
In the noninteracting case, we can see the low-temperature upturn
of the nearest-neighbor magnetization, as discussed in Ref.
\cite{JChang:2004} (note at $T\rightarrow 0$, the noninteracting
local susceptibility always $\rightarrow 0$ since the impurity
resonance sits at a finite energy for any generic potential).  On
the other hand, the upturn of the magnetization of the sites
contributing to the linewidth is much weaker to nonexistent in the
noninteracting case, although the LDOS effect still manifests
itself via the fact that $\Delta \nu_L < \Delta\nu_H$.  In the
$U>0$ case, however, the upturns are much stronger,  and even
manifest the saturation of the magnetization arising from $\Delta
\nu_L$ sites observed in experiment \cite{SOuazi:2006}, which
leads to the surprising increase of the $\Delta \nu_H-\Delta\nu_L$
observed.

\begin{figure}[t]
\begin{center}
\leavevmode
\includegraphics[clip=true,width=.9\columnwidth]{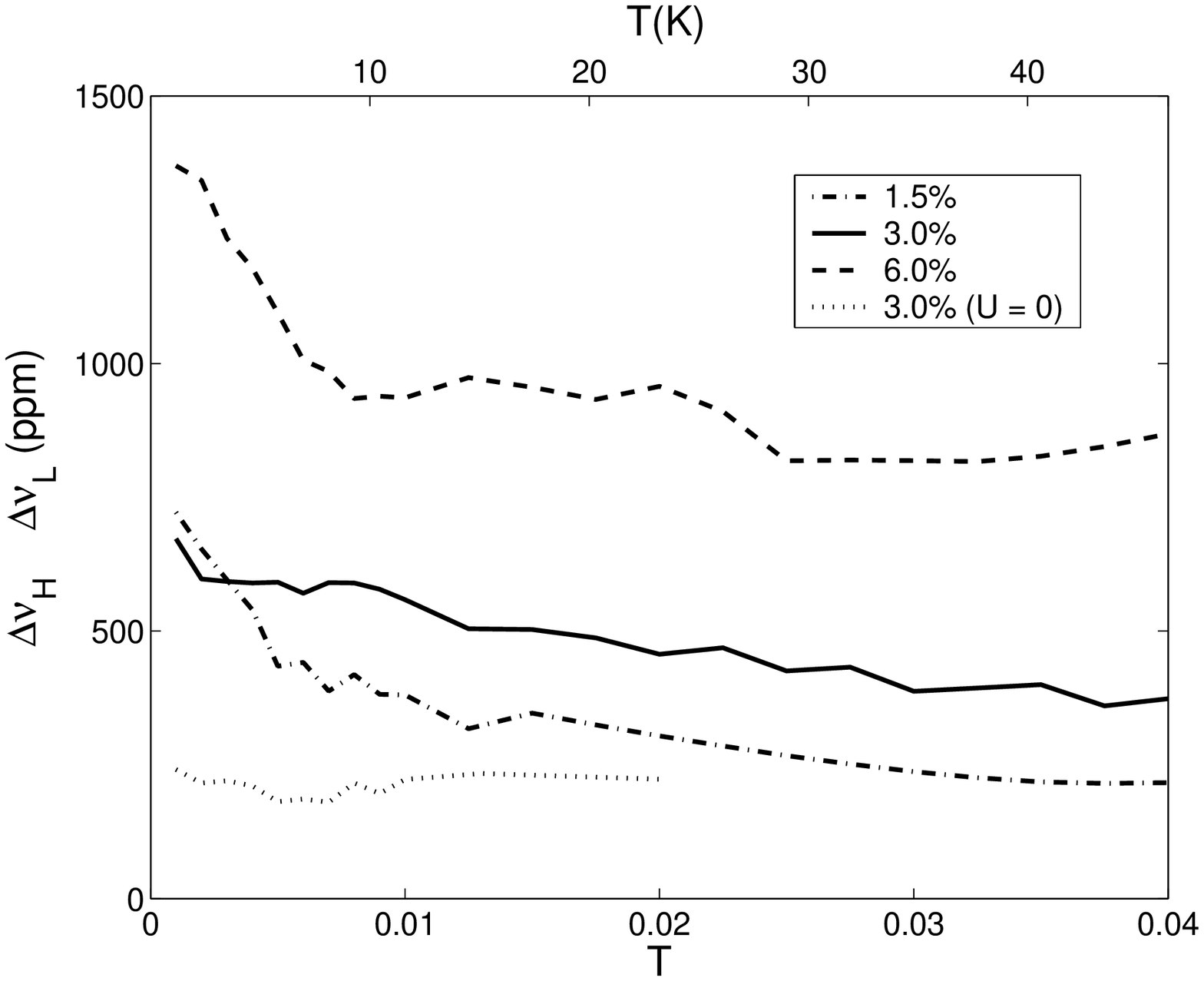}
\includegraphics[clip=true,width=.9\columnwidth]{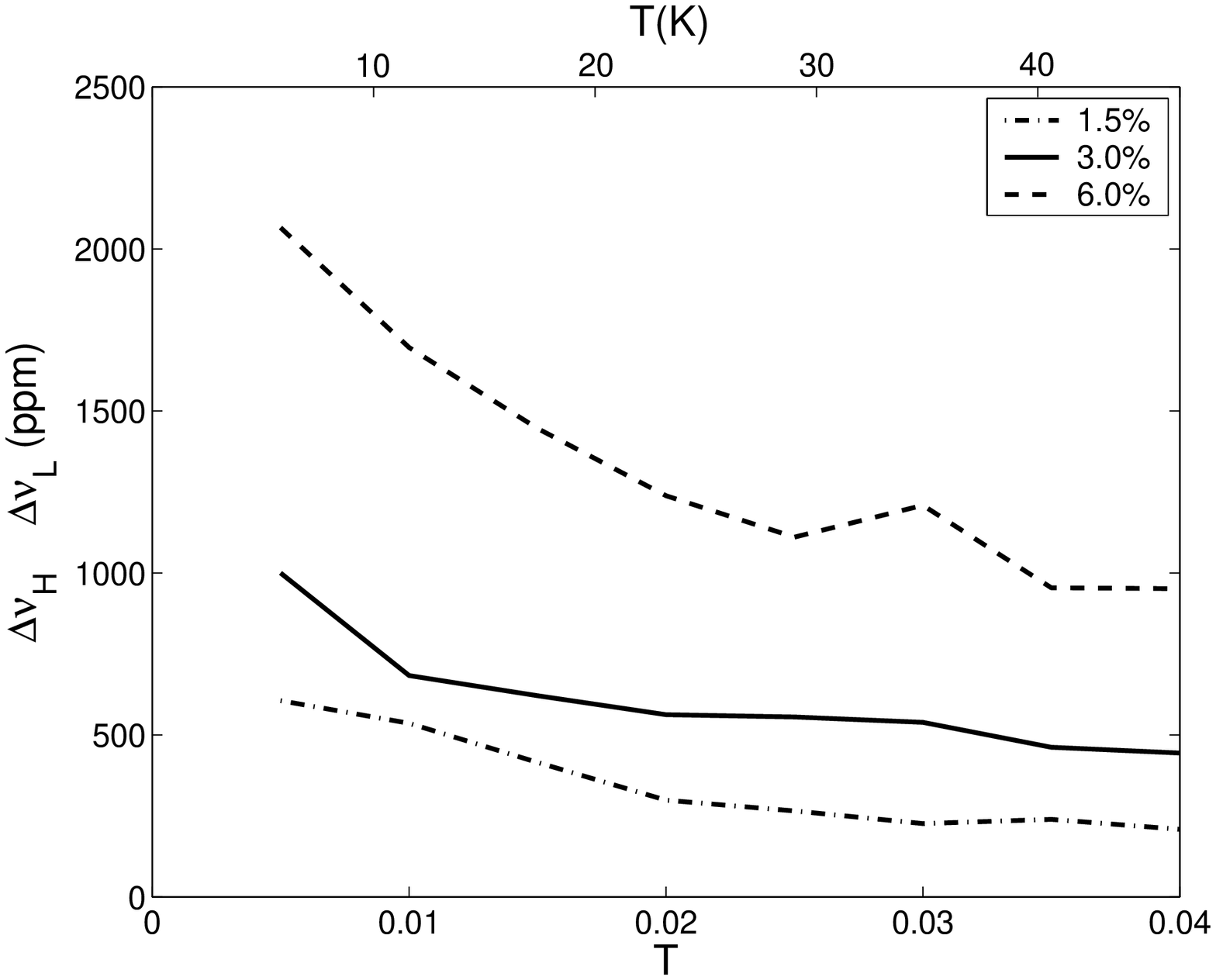}
\caption{Line asymmetry--theory: difference of halfwidths on
negative ($\Delta \nu_L$) and positive ($\Delta \nu _H$) frequency
sides of NMR lines vs. $T$ for five  impurity concentrations. Top
panel: $U=1.75$, $g\mu_B B/2$=0.004, $V_{imp}$=100.  The result
for $U$=0 is also shown for comparison (dotted line).   Bottom
panel: same, but with $V_{imp}=10$. } \label{fig:lowTupturns}
\end{center}
\end{figure}

\begin{figure}[t]
\begin{center}
\leavevmode
\includegraphics[clip=true,width=.9\columnwidth]{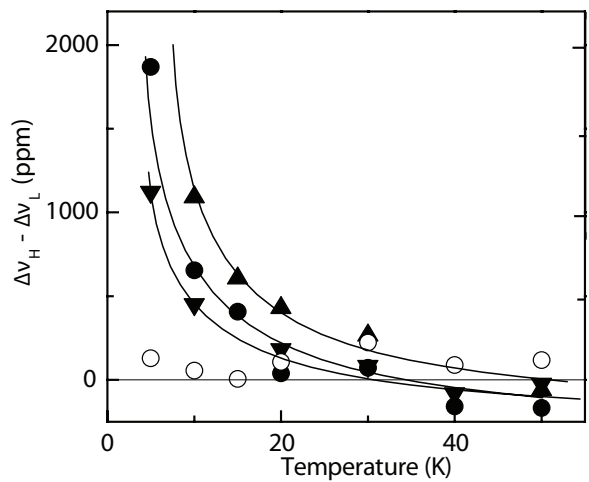}
\caption{Line asymmetry--experiment: difference of halfwidths on
negative ($\Delta \nu_L$) and positive ($\Delta \nu _H$) frequency
sides of NMR lines vs. $T$ for three impurity concentrations, from
Ref. \onlinecite{SOuazi:2006}.} \label{fig:lowTupturnsexpt}
\end{center}
\end{figure}

To verify if the phenomenon is reproduced quantitatively, we plot
explicitly the difference of the half width at half-maximum on the
high frequency side of the $^{17}O$ line, $\Delta \nu_H$, and the
same quantity on the lower side $\Delta \nu_L$.  Each is
independently enhanced at low temperatures, but it is the
difference which is particularly striking, as seen in Fig.
\ref{fig:lowTupturns}, where theory for $U=1.75$ is compared with
the Ouazi {\sl et al.}\cite{SOuazi:2006} results shown in Fig.
\ref{fig:lowTupturnsexpt}. While the details of the theoretical
curves do not agree exactly with experiment, it is clear that the
basic results are reproduced by the theory, both in terms of the
temperatures at which the upturns begin, and in terms of the
magnitudes of the upturns themselves.  On the other hand, the
lower halfwidth $\Delta \nu_L$, is roughly $T$ independent in
experiment, but has a weaker but still significant enhancement in
the calculation, as indicated  in Fig. \ref{fig:key_sites}. We do
not understand the origin of this discrepancy at present. In
addition, the theoretical result retains a certain asymmetry of
the lineshape up to higher temperatures, whereas the experimental
lineshape becomes symmetric above about 40K.  This may be due to
the neglect of inelastic scattering, which becomes important at
higher $T$, in the calculation.

The two figures shown in Fig. \ref{fig:lowTupturns} correspond to
two different values of the impurity potential $V_{imp}$.  The
upper panel corresponds to a value $V_{imp}=100$, which gives an
impurity resonance energy of $\Omega_0 \simeq -0.01t$, whereas the
$V_{imp}=10$ lower panel corresponds to a resonance energy of
$\Omega_0 \simeq 0$ within our numerical resolution, for this
particular band. Thus the local Fermi level density of states near
the impurity is larger in the second case, leading indeed to a
stronger upturn, as anticipated. The actual resonance energy of a
Zn in YBCO is not known at this time, but is expected to be close
to the -1.5meV observed in BSCCO-2212.

\section{Further comparisons with experiment: $^7$Li } \label{Li}
The ``universality" of the magnetic response to strong in-plane
{\it nonmagnetic} defects was noted early on by Bobroff {\sl et
al.} \cite{JBobroff:1999}.  That is, any in-plane impurity which
from a conventional chemistry standpoint is expected to be
nonmagnetic appears to have a nearly identical effect on both
normal state and superconducting properties.  This includes Zn, Li
and defects in the plane created by electron irradiation, which
produce nearly identical changes in susceptibility, $T_c$, and
resistivity per in-plane impurity. This is remarkable because
there would appear to be important electronic structure
differences between the Zn ion, which has a closed shell, and a
Li, which is believed to localize a hole around itself.  It is
generally believed that the essential features of these defects
are therefore simply their ability to exclude mobile conduction
electrons, hence our choice of model of, e.g. Zn as a strong
repulsive potential.  Within this assumption of ``universality" of
in-plane nonmagnetic defects, we can take the results for the
magnetization distribution in the disordered system already
produced, and use them to describe the results of earlier $^7$Li
NMR experiments on YBCO.

\begin{figure}[t]
\begin{center}
\leavevmode
\includegraphics[clip=true,width=.99\columnwidth]{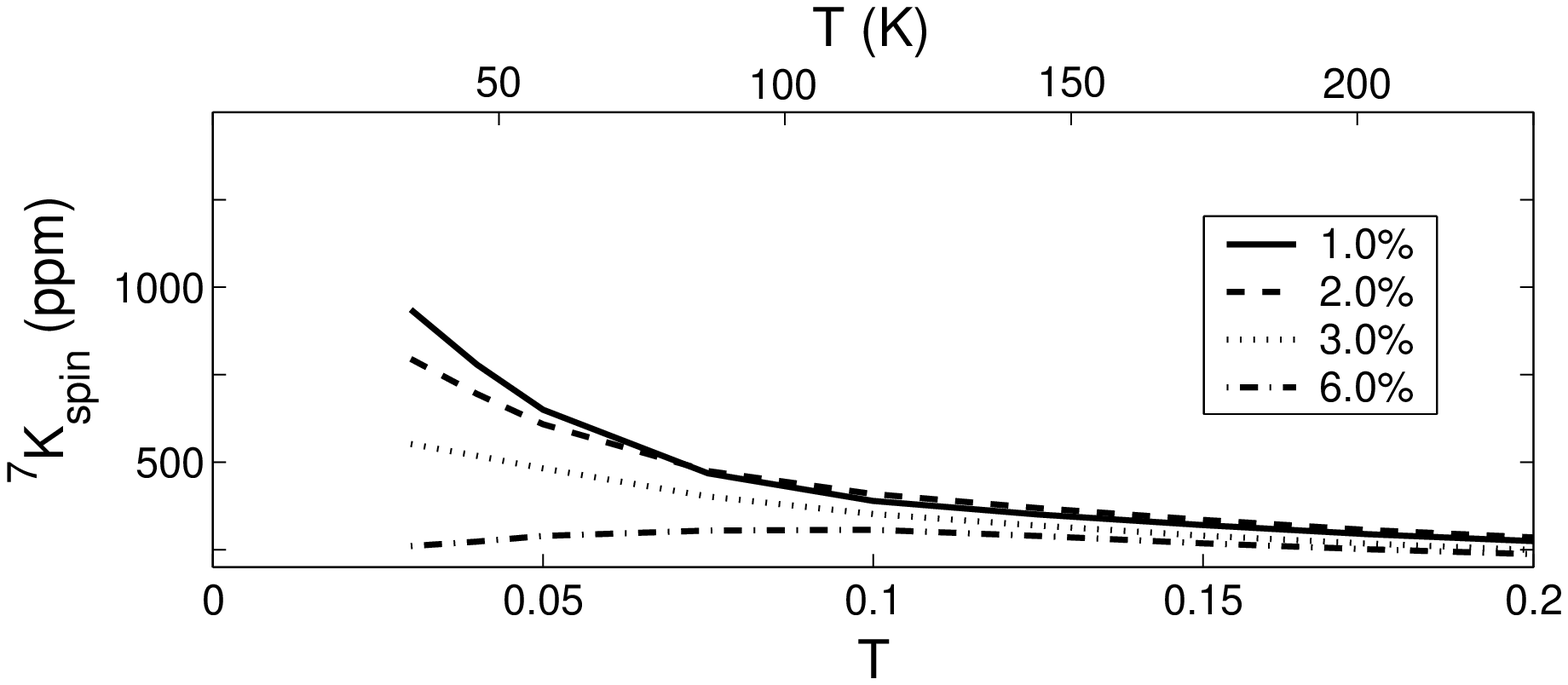}
\includegraphics[clip=true,width=.99\columnwidth]{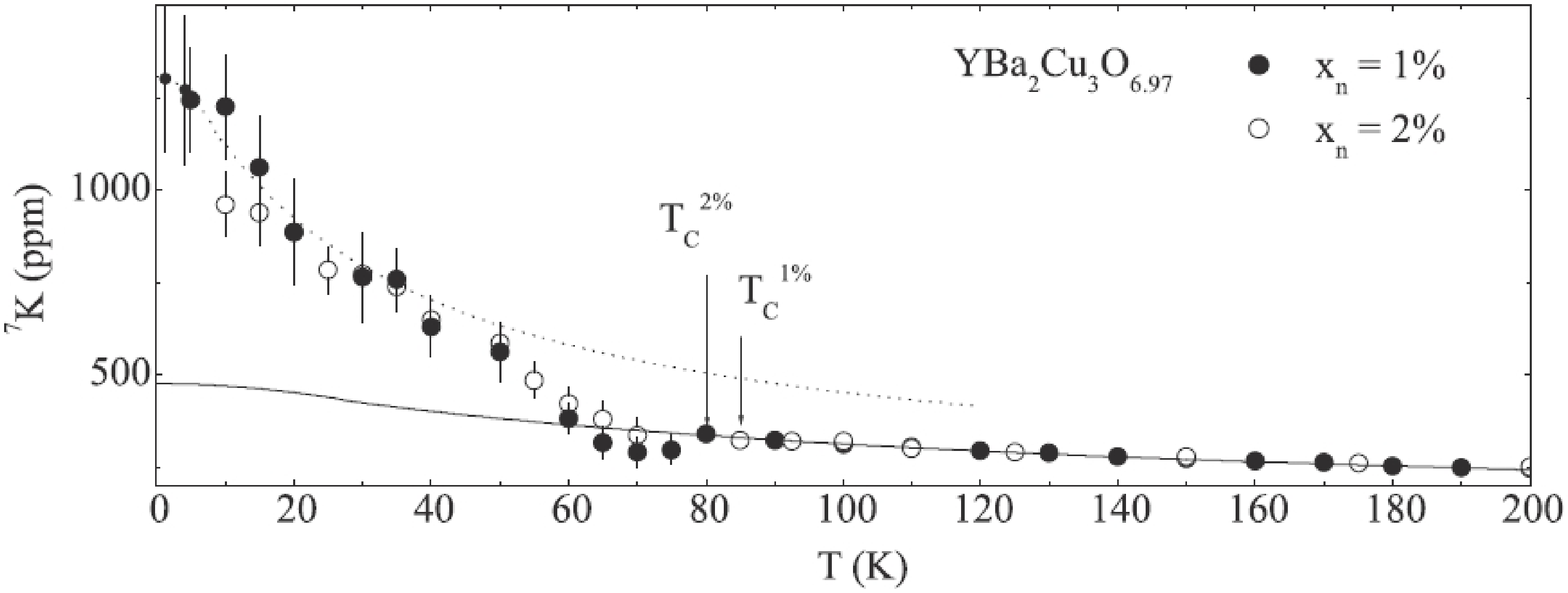}
\caption{Upper panel: Li Knight shift as calculated in text, using
same magnetizations, obtained for same parameters as shown in the
histograms in Fig. \ref{fig:impMags}, 3 random configurations.
 Lower panel: experimental Knight
shift data from Ref. \onlinecite{JBobroff:2001} for $x_p$= 1\% and
2\% Li in optimally doped YBCO for $B$=7T.} \label{fig:LiShiftvsT}
\end{center}
\end{figure}

Li has the thus far unique ability in the cuprates to
simultaneously provide an in-plane impurity and a nucleus ($^7$Li)
suitable for NMR.  The signal is therefore not complicated by
contributions from regions of the sample far from impurity sites,
but provides direct information about the immediate vicinity of
the impurity, which replaces a Cu in the CuO$_2$ plane.  We will
assume, as in prior work, that the Li provides a shift equal to
the sum of the magnetizations on its nearest neighbor sites,
leading to a shift (compare Eq. (\ref{localKnight})).

\begin{equation}
^{7}K_{spin}(\rr)=\frac{^{7}A_{hf}\sum_\delta \chi
_{spin}(\rr+\delta)  }{\mu _{B}},
\end{equation}
where $\delta$ is a nearest neighbor displacement.  Results plotted
as a function of temperature are shown in Fig. \ref{fig:LiShiftvsT}
and compared to the experimental results on Li substituted in
optimally doped YBCO of Bobroff {\sl et al.} \cite{JBobroff:2001}.
The theory seems to reproduce the initial weak dependence on the Li
concentration (nonexistent in the normal state, small in the
superconducting state).  This is because the Li are sensitive to the
local environment of the impurity only, and the shift does not
therefore depend on concentration until interference effects become
significant.  Until now, experiments have only been performed on a
maximum of 2\% planar Li impurities; the deviations of the curves
corresponding to higher concentrations at low temperatures
constitute predictions of the theory which can be verified by NMR.
We note that the experimental results, in particular the change in
slope near $T_c$ and the subsequent rise at low $T$, have been
discussed in terms of the effect of the opening of the
superconducting gap on Kondo screening of simple moments by the
$d$-wave quasiparticle gas\cite{Fradkin}.  Our results here, which
do not account for the spin-flip scattering necessary for the Kondo
effect to take place at all,  indicate that a strong-correlation
explanation for these, and associated phenomena, is more likely.

\section{Conclusions}

We have shown that a theoretical model of strong impurities in a
$d$-wave superconductor, with residual quasiparticle interactions
treated in a  weak coupling Hubbard model within a mean-field
approximation, provides an excellent description of NMR
experiments on optimally doped YBCO.  This model assumes that both
Zn or Li impurities act as simple strong potential scatterers
which induce staggered magnetic polarization clouds around
themselves with very small  total moment {\it proportional to the
external magnetic field}. The results and comparison with
experiments on optimally doped YBCO are {\it not} compatible with
static impurity-induced magnetism present in zero field.  This is
consistent with the lack of a magnetic $\mu$SR signal over most of
the YBCO phase diagram down to very low
doping\cite{SonierRMP,Sonier}, and with the absence of a Schottky
anomaly in specific heat measurements\cite{DLSisson:2000}. The
universal response of the system to nonmagnetic defects which are
very different chemically also confirms the primacy of the strong
correlations in the CuO$_2$ plane for these phenomena.

The ability of NMR to probe nuclei with  different configurations
with respect to the Cu spins means that different interesting
aspects of the physics can be tested by the model.  In particular,
the $^{17}$O line is primarily a probe of the weaker magnetism
induced far from the impurities.  It is thus more sensitive to the
effects of interference of the quasiparticle bound states
associated with  different scattering centers.  We have shown that
this quasiparticle interference is responsible for the enhancement
of the $^{17}$O linewidth as measured by experiment, and the
variation of this lineshape with Zn concentration  has been
reproduced quantitatively by the model.  The low-temperature
enhancement of the $^{17}$O line to the positive side caused by Zn
appears, within this framework, to be related to the LDOS
enhancement around the impurities, as suggested by Ouazi {\sl et
al.}\cite{SOuazi:2006}.  On the other hand, the size and strong
temperature dependence of the line broadening can be understood
only if electronic correlations and concomitant paramagnetic local
moments are present.

The $^7$Li nucleus, which senses only the nearest neighbor Cu,
gives us a picture into magnetic effects in the vicinity of the
individual impurities.  Our study, which reproduces quantitatively
the $^7$Li Knight shift magnitude and $T$ dependence in both the
normal and superconducting states,   concludes that experiments on
Li done thus far have probed only the regime of individual
isolated impurities, where interference effects are not strong
enough to affect nearest neighbor sites, but these should be
visible by going to only slightly higher Li concentrations.  In
addition, it shows how resonant states in the $d$-wave
superconductor  enhance both the single-impurity magnetic effects
and the quasiparticle interference.

We note a number of questions which are open and which should be
clarified in subsequent work.  First, the current framework is a
weak-coupling mean field approach which neglects the
renormalization of the electronic structure near the impurity
sites found in strong-coupling approaches.  We anticipate that
such effects, which arise from diagrams not included in the
RPA-type analysis here, will give rise to quantitative changes in
the values of parameters chosen here to fit experiment, but not
alter the overall physical picture we have presented.  Many of
these issues have already been raised and understood in the
context of homogeneous RPA-level spin-fluctuation theories, where
a reduced  effective interaction $U$ replaces the bare $U$ after
resummation of a subclass of diagrams.  It is intriguing to note
that NMR experiments seem to require us to work very close to the
magnetic phase transition in the theory in order to explain e.g.
the magnitude of the linewidth and the $T$ dependence of the line
asymmetries.  This is reminiscent of the early fits of spin
fluctuation theories of NMR for the homogeneous systems, which
 required values of the effective interaction to be close to
the bulk transition in the model.  Taken together, it is tempting
to speculate that these results indicate the flow of the true
Hamiltonian of the system to strong coupling at sufficiently low
energies.

In this work we have not investigated carefully the transition to
the normal state, which we reserve for a subsequent paper.  It is
interesting to note that, while earlier works on the normal state
using the same model\cite{NBulut:2001,YOhashi:2001} found it
necessary to introduce an extended impurity potential to fit
experiment, our conclusion is that this is not necessary in the
superconducting state. Following Bulut's
suggestion\cite{NBulut:2003} that the extended nature of
scattering potential may arise in the normal state from the
renormalization of the local electronic structure by correlations,
this may be an indication that the processes leading to this
effect are simply gapped below $T_c$.

Finally, we emphasize that our calculation does not include
spin-flip scattering terms necessary to recover a true Kondo
effect, and that to the extent our analysis has been successful,
our results therefore imply that Kondo physics is probably
unnecessary to describe the phenomena in question, at least at
optimal doping. It will be interesting to see whether the
phenomenology put forward here continues to hold as one goes to
underdoped systems with planar impurities which remain
paramagnetic.  Other possible extensions include more strongly
interacting and/or more disordered systems, where static magnetism
in zero field may be present, studies of the normal state, as well
as of the overdoped regime.  Work along these lines is in
progress.

\label{conclude}

~

{\it Acknowledgements.} The authors acknowledge stimulating
conversations with H. Alloul and T. Nunner.   Partial support for
this research (B. M. A. and P. J. H.) was provided by ONR
N00014-04-0060, and by a visiting scholar grant from C.N.R.S.
(P.J.H.).

\end{document}